\documentclass[aps,prb,twocolumn,preprintnumbers,superscriptaddress]{revtex4-2}
\usepackage{latexsym,epsfig,amsfonts,color,bm,times,hyperref,amstext,epstopdf,float}
\usepackage{blindtext}
\usepackage{amssymb,array}
\usepackage{physics}
\usepackage[utf8]{inputenc}
\usepackage{graphicx}
\usepackage{amsmath, amssymb}
\usepackage{bbold}
\usepackage{braket}
\usepackage{xcolor}
\usepackage{wrapfig,dsfont,stmaryrd}
\usepackage{wasysym}
\usepackage{wrapfig}

\graphicspath{{entanglement_figures/}}

\begin{document}
	
\title{Persistent entanglement of valley exciton qubits in transition metal dichalcogenides integrated into a bimodal optical cavity}
	%\author{Fanyao Qu}
	%\thanks{fanyao@unb.br}
	%\affiliation{Instituto de F\'{\i}sica, Universidade de %Bras\'{\i}lia, Bras\'{\i}lia-DF 70919-970, Brazil}
	%\author{Lorena Reis de Lima}
	%\thanks{limalreis@gmail.com}

\affiliation{Instituto Federal do Triângulo Mineiro, Patrocínio MG, 38747-792, Brazil}
\affiliation{Instituto de F\'{\i}sica, Universidade de Bras\'{\i}lia, Bras\'{\i}lia-DF 70919-970, Brazil}
\affiliation{School of Physics, State Key Laboratory of Crystal Materials, Shandong University, Jinan, China}
\affiliation{International Center for Condensed Matter Physics, University of Bras\'{\i}lia, 70910-900, Bras\'{i}lia, Brazil}
\affiliation{Instituto de F\'{\i}sica, Universidade Federal de Uberlândia, 38400-902 Uberlândia, MG, Brazil}

\author{Borges H. S.$^{1}$, Celso A.N. J\'{u}nior$^{2}$, David S. Brandão$^{2}$, Fujun Liu$^{2}$, V. V. R. Pereira $^{2}$, S.J. Xie$^{3}$}

\author{Fanyao Qu$^{1,4,\ddagger}$}
\author{ A.M. Alcalde$^{5,\ddagger}$}
\date{\today}
	
	\begin{abstract}
		
		 We report dissipative dynamics of two valley excitons residing in the $K$ and $K^\prime$-valleys of bare WSe$_2$ monolayer and the one being integrated into a bimodal optical cavity. In the former, only when the exciton-field detunings in the $K$ and $K^\prime$-valleys are rigorously equal (resonant detuning), partially entangled stationary states can be created. Otherwise the concurrence of exciton qubits turns to zero. Remarkably, in the latter (the WSe$_2$ monolayer in a bimodal optical cavity), the transfers of entanglement from one subsystem (exciton/light) to the other (light/exciton) take place. Hence a finite stationary concurrence of exciton qubits is always generated, independent of whether the exciton-field detuning in two valleys is resonant or non-resonant. In addition, it can even reach as high as 1 (maximally entangled state of two valley excitons). Since there no real system which has a strictly resonant detuning, an immersion of the WSe$_2$ monolayer in a bimodal optical cavity provides an opportunity to overcome the challenge facing by the bare WSe$_2$, opening a novel realm of potential qubits. 
	
		 %Furthermore the lifting of the excitonic state degeneracy suppresses stationary concurrence, while it can be partially compensated by an increase of electron-hole exchange interaction.

	\end{abstract}
	
	\maketitle
	
	\section{introduction}
	
\begin{figure}[H]
		\centering
		\includegraphics[width=\linewidth]{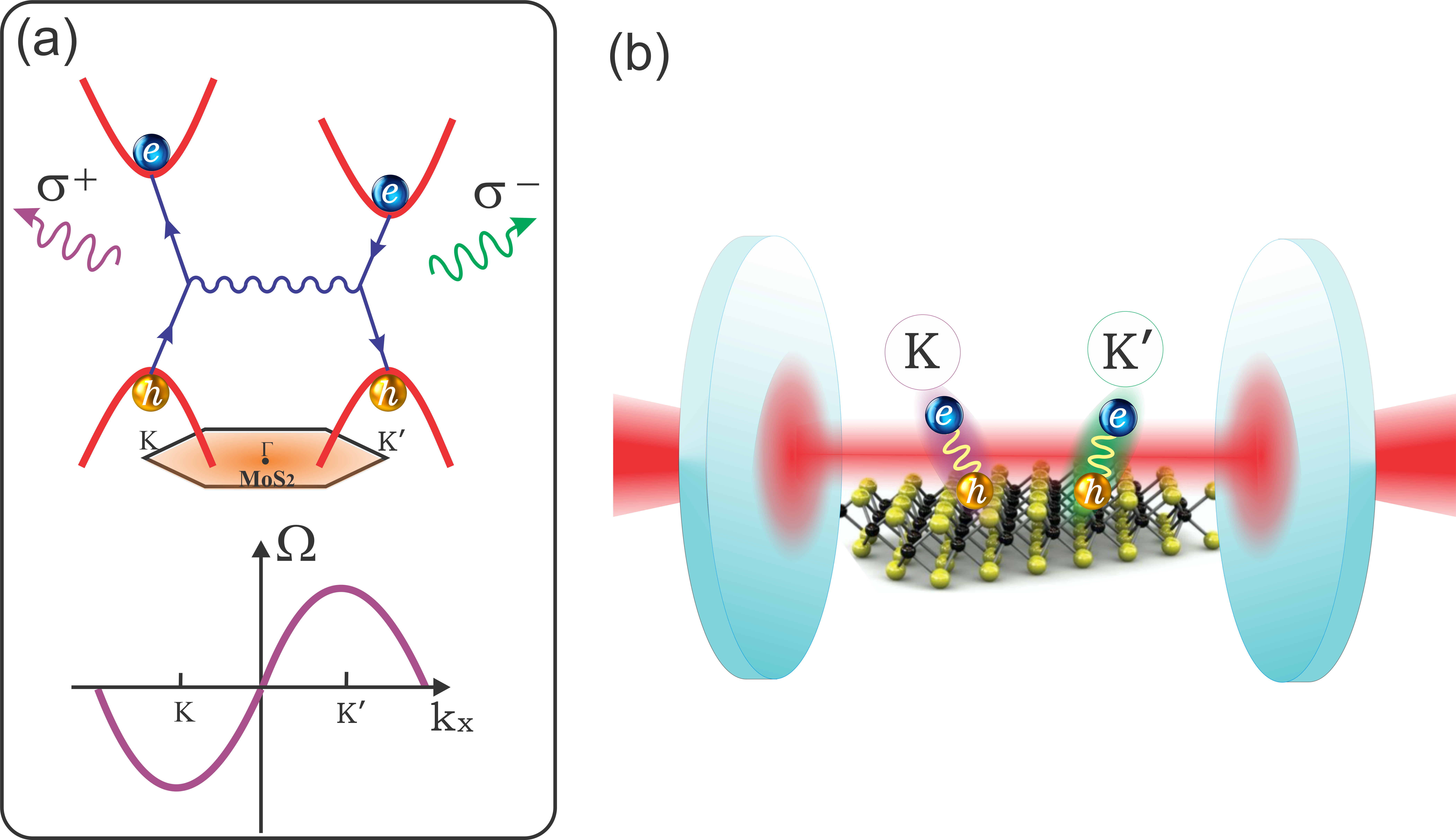}
		\caption{(a) Upper part: Schematic diagram of Brillouin zone (filled hexagonal),  valley excitons formation and Feynman diagram of EXI (lines with arrows); lower part: Berry curvature of the lowest conduction band in K and K$^\prime$ valleys, and (b) TMD integrated nanocavity system.}
		\label{Figure2}
	\end{figure}

	Monolayer transition metal dichalcogenides (TMDs), denoted by MX$_2$ with M=Mo,W and X=S,Se, possess a broken inversion symmetry which leads to a direct band gap at two degenerate but in equivalent valleys $K$ and $K^{\prime}$. Although these two states are energetically degenerate, the orbital angular momentum quantum number of the valence band (VB) in $K$ valley is +2 whereas it is -2 in $K^{\prime}$ valley. Moreover, the strong spin-orbit coupling (SOC) splits the band edge states, resulting in an energy separation between spin-up and spin-down states reaching 0.45 eV in the VB and tens of meV in the conduction band (CB) of monolayer WX$_2$. In addition, time reversal symmetry preservation of the TMDs requires that the spin splitting in different valleys must be opposite. As a result, the spin-down (spin-up) state in $K$ valley would have the same energy with its corresponding spin-up (spin-down) state in $K^{\prime}$ valley, see Fig.\ref{Figure2}. These characteristics of the monolayer TMDs lead to so-called spin–valley locking and valley dependent optical selection rules.
	In turn, when the electrons recombine to the holes in the $K$ ($K^{\prime}$) valley, they can only generate left (right) circularly polarized light. Furthermore, because of spin–valley locking, the intervalley scattering happens only when the intervalley spin flips and the large momentum transfer ($\approx K - K^{\prime}$) simultaneously occur. This strongly suppresses the intervalley scatterings, endowing a long valley polarization time of the carriers. Recently, quantum coherence of valleys has been experimentally demonstrated,~\cite{Jones2013}, coherent control of valley-spin qubits has also been theoretically proposed \cite{Brooks2020,Szechenyi2018}. Therefore the valley of the monolayer TMDs has potential to be used as quantum information carriers. 
	
	Owing to the two-dimensional (2D) spatial confinement and reduced dielectric screening, the monolayer TMDs exhibit strong Coulomb interactions \cite{xiao2012coupled,Wu2015,Dias_085406_2020}. Under light excitation, the formation of tightly bound electron-hole pairs, so-called excitons, is favored~\cite{Wu2015,qu2019controlling,Tang2019,Dias_085406_2020}. Owing to optical valley selectivity, namely, the right-handed (left-handed) circularly polarized light only creates the excitons in $K (K^{\prime}$) valley, the excitons will be endowed with a valley degree of freedom, denominated as valley excitons. In addition, their high binding energies of up to a few hundred meV make excitons quite stable, even at room temperature~\cite{Dias_085406_2020,schaibley2016valleytronics,wang2020spin}. It sheds light on room temperature valley exciton qubits.
	
	The ﬁeld of quantum computing was ﬁrst foreseen by Richard Feynman in the early 1980s with the aim of solving quantum mechanics problems. Since then, the interest of both fundamental-science researchers and industrial-actors like Google, IBM, Intel, and Lockheed Martin, has increased exponentially in this field. 
	Many qubits, the fundamental building block of a quantum computer, such as electronic spin, nuclear spin, electronic states of trapped ions, superconducting qubit and topological qubit, have been proposed and explored. Although a great progress in quantum computing has been made such as processing velocity has grown exponentially in a recent few years, it is still facing  high challenges. For example, the ion qubit needs to be in a high vacuum, the superconducting qubit has a short decoherence time and must be kept very cold (below 100 mK, or 0.1 degrees above absolute zero), etc. To shed light on overcoming these challenges, we propose another alternative: valley exciton qubits implemented in TMDs. The large thermal stability, long valley coherence times combining with fast and conveniently operation by circularly polarized laser field make valley exciton a great compromising quantum information carrier. 
	
	In this work, we focus on the study of quantum entanglement of inter-valley excitonic states, special interest is devoted to the optical generation of entanglement and the injection of excitonic entanglement into the system and its availability over long time periods for applications in quantum information processing. The concurrence which is used to measure entanglement of valley excitons, however, always falls down to a vanishing small value for a non-resonant detuning (exciton-ﬁeld detuning in valley $K$ is different with that in $K^{\prime}$). Interestingly, the situation is changed for the TMDs integrated into a bimodal optical cavity. By tuning the initial state of the system, exciton-ﬁeld detunings, electron-hole exchange interaction (EXI) and the coupling between exciton and cavity mode, the concurrence of two exciton qubits can remarkably reach as high as 1. It is attributed to cavity enhanced light-matter interaction. These results demonstrate the viability and tunability of valley exciton qubits which possess high-speed operation, easy and convenient control, long coherence time and capability of working at room temperature.

	\noindent	
	
%\textcolor{red}{Quantum entanglement, in the quantum information scenario, is considered a necessary resource to implement non-local quantum operations between two or more agents [referencias]. 
%The experimental implementation of the applications of quantum entanglement, such as quantum teleportation, quantum information, superdense coding, is challenging and a constant exploration of physical systems that allow the construction and preservation of entangled quantum channels is necessary.}

\section{Theoretical Framework}

\begin{figure}[H]
		\centering
		\includegraphics[width=\linewidth]{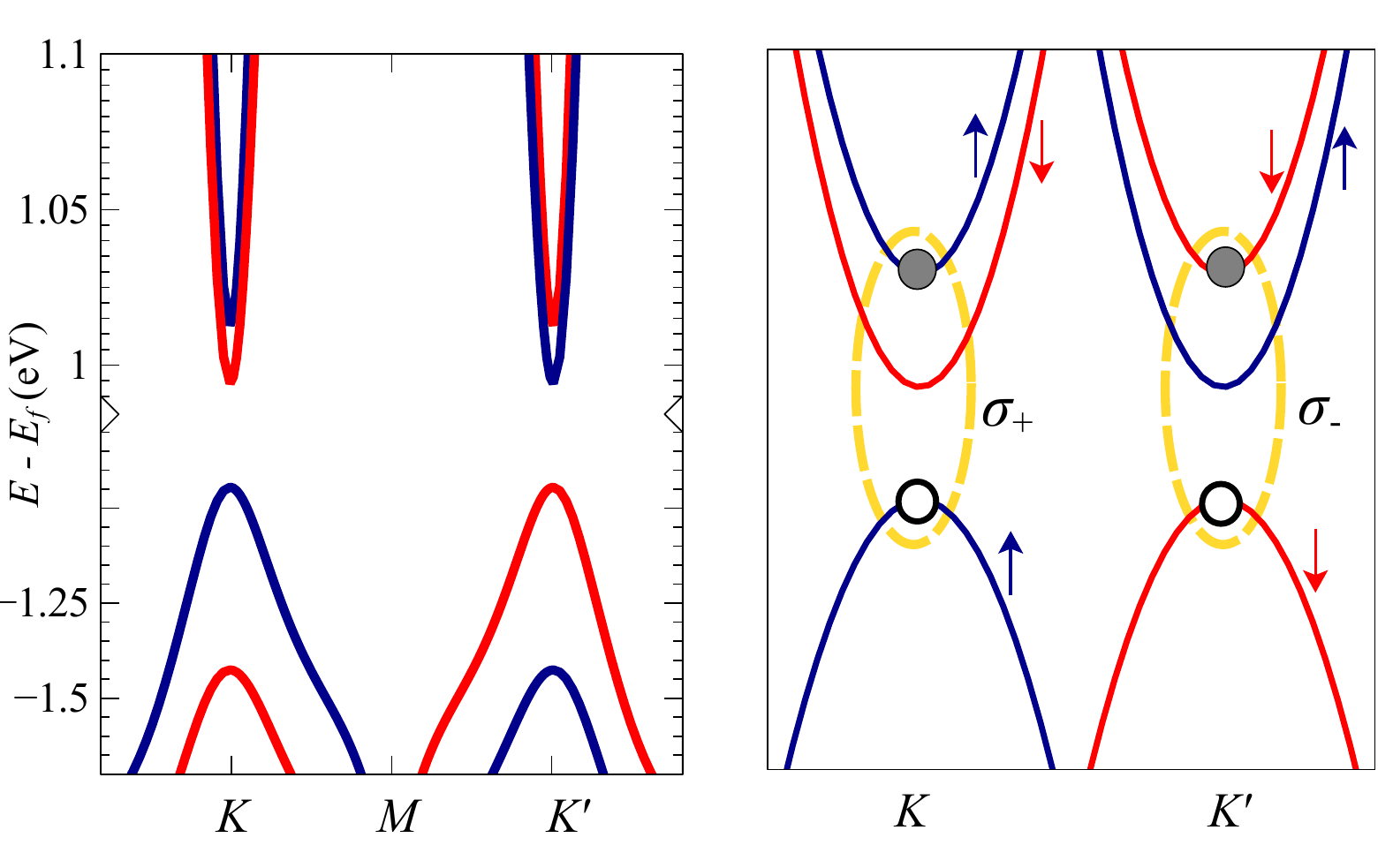}
		\caption{(a) Band structure of monolayer WSe2 nearby $K$ and $K^{\prime}$ points, and (b) schematic representation of valley bright exciton formations. Red and blue curves correspond to spin-up and spin-down states, respectively. }
		\label{Figure2b}
	\end{figure}

Here we chose as physical system for quantum entanglement implementation the tungsten diselenide (WSe$_2$) . The band structures and wavefunction of WSe$_{2}$ are obtained by density functional theory (DFT) calculations using the QUANTUM ESPRESSO package~\cite{Giannozzi2009} (see Figure \ref{Figure2b}). The kinetic energies cutoff for wave function (ecutwfc) and charge density (ecutrho) are set as 600 and 60 Ry, respectively. The range-separated hybrid functional proposed by Heyd, Scuseria, and Ernzerhof (HSE06)~\cite{Heyd2003,Krukau2006} is adopted for the exchange-correlation energy. Numerical integrations in the Brillouin zone are evaluated with the Monkhorst-Pack mesh of $12\times12\times1$. All structures are relaxed until the total energy converges to within $10^{-4}$ eV during the self-consistent loop, employing the Methfessele-Paxton method. To correctly account for the strong spin-orbit coupling arising from heavy tungsten atoms in WSe$_2$ \cite{Tang_14529_2019,Dias_085406_2020}, the spin-orbit coupling corrections were included in the electronic property calculations.

With the obtained energies and wavefunctions, we develop a tight-binding (TB) model using the package Wannier90~\cite{Mostofi2014}. This maps the ground-state wave functions from the density functional theory output file onto a maximally localized Wannier function basis $\{\vert{W}_{i\mathbf{R}}\rangle\}$, where $i=(I, \alpha)$ is the composite index of the atomic orbit $\alpha$ of a atom with atom site $\mathbf{r}_I$. $\mathbf{R}$ = $R_{a}\mathbf{a}$ + $R_{b}\mathbf{b}$ + $R_{c}\mathbf{c}$ is the Bravais lattice vector with $R_{a/b/c}$ being the Bravais lattice vector component on the direction of the unit cell lattice vector $\mathbf{a}/\mathbf{b}/\mathbf{c}$. The basis function of TB model is thus given by
\begin{equation}
\vert\phi_{i\mathbf{k}}\rangle = \frac{1}{\sqrt{N}}\sum_{\mathbf{R}}\,e^{-i\mathbf{k}\cdot\mathbf{R}}\vert{W}_{i\mathbf{R}}\rangle,
\label{wannierfunction}
\end{equation}
and elements of the $TB$ Hamiltonian $\mathbf{H}^{TB}$ are given by
\begin{equation}
 \mathbf{H}^{TB}_{ij}(\mathbf{k}) = \sum_{\mathbf{R}}e^{-i\mathbf{k}\cdot\mathbf{R}}\,t_{ij}(\mathbf{R}),
\end{equation}
where $t_{ij}(\mathbf{R})$ = $\langle{W_{i0}}\vert{H_{KS}}\vert{W}_{j\mathbf{R}}\rangle$ is the matrix element extracted from the Wannier90 output file, with $H_{KS}$ being the Hamiltonian of Kohn-Sham equation.

	\noindent

	\section{Valley exciton in WSe$_2$ monolayer}
	
	\subsection{Bethe–Salpeter equation for valley Exciton}
	\label{sectime}
	
	With eigenvalues and wave functions of the single particle at hand, we can calculate excitonic states through many-body problem using {BSE}. The exciton Hamiltonian is composed of the electron $H_{e}$ and the hole $H_{h}$ single particle Hamiltonians plus Coulomb interaction $V_{\vec{R}}$ which couples the electron-hole pairs
	\begin{equation}
		H_{X}= H_{e}+H_{h}+V_{\mathbf{R}}.
	\end{equation}
	
	Due to the finite width of TMD monolayer and the spatial inhomogeneity of the dielectric screening environment, we choose to modeling the interaction Coulomb potential, using Keldysh formalism~\cite{Wu2015}
	\begin{equation}
		V_{\mathbf{R}}= -\frac{e^{2}}{8 \epsilon_{0} \epsilon_{d} r_{0}} \left[ H_{0} \left( \frac{|\mathbf{R}|}{r_{0}}\right) -Y_{0} \left( \frac{|\mathbf{R}|}{r_{0}}\right) \right],
	\end{equation}
	where $H_{0}$ and $Y_{0}$ are the Struve and Bessel functions of the second kind, respectively; $\epsilon_{d}$ is the effective dielectric constant ($\epsilon_{d}=2.39$ for a SiO$_2$ substrate); and $r_{0}$ represents a characteristic length, defined by $r_{0} = 2 \pi \chi_{2D}/\epsilon_{d}$, with $\chi_{2D}$ being the two-dimensional polarizability.
	
	%{\color{red}VAMOS MANTER ESSA PARTE RELACIONADA AOS TIPOS DE SUBSTRATO? SERA QUE OS REFEREES NÃO VÃO ACHAR QUE NÃO ESTA RELACIONADO COM O RESTO DO TRABALHO, FOCADO EM TRION? In this work we consider three types of substrate, vacuum above and below monolayer ($\epsilon_{d}=1$) , vacuum above and SiO$_2$ below ($\epsilon_{d}=2.39$) and the monolayer encapsulated in hBN ($\epsilon_{d}=4.4$). The effective dielectric constant given by $\epsilon_{d}=(\epsilon_{above}+\epsilon_{below})/2$.} Furthermore, \hh{the dielectric constant is taken to be $\epsilon_{d}=2.39$ (SiO$_2$ substrate),} $r_{0}$ represents a characteristic length defined by $r_{0}=2 \pi \chi_{2D}/\epsilon_{d}$, with $\chi_{2D}$ being the two-dimensional polarizability.
	
	Since we study periodic lattices, it is more convenient to work in the reciprocal space; in $k$-space the Keldysh potential acquires the form~\cite{Wu2015,Dias_085406_2020}
	\begin{equation}
		V_{\mathbf{q}}=-\frac{e^{2}}{2 \epsilon_{0} \epsilon_{d} |\mathbf{q}| (1+r_{0} |\mathbf{q}|)},
	\end{equation}
	where $\mathbf{q}=\mathbf{k}-\mathbf{k}^{\prime}$.  It is worth noting that a controllable tuning of the band gap and the exciton binding energy can be realized by engineering the dielectric environment of the monolayer, e.g., by encapsulating the TMDC between other layered materials, or modifying the substrate. 
	
An exciton state $|\Psi_{S}(\mathbf{Q}) \rangle$ is a  coherent superposition of hole (with crystal momentum $\mathbf{k}$) and electron (with crystal momentum $\mathbf{k}+\mathbf{Q}$) states from band-pairs ($v$, $c$) in the reciprocal space and can be written as

	\begin{equation}
		|\Psi_{S}(\mathbf{Q}) \rangle= \sum_{c,v,\mathbf{k}} A_{c,v,\mathbf{k},\mathbf{Q}}^{S}  |c,\mathbf{k}+\mathbf{Q} \rangle \otimes |v,\mathbf{k} \rangle   , 
		\label{Exc_Basis}
	\end{equation}
	where $S$ and $\mathbf{Q}$ are the exciton band index and the center of mass momentum, respectively. In this study, we focus on the low energy excitons. Thus we only consider the two highest VBs and the two lowest CBs.
	The exciton eigenvalue problem in the basis displayed in Eq.~(\ref{Exc_Basis}) leads to the following the Bethe–Salpeter equation,	
	\begin{align}
		E_{S}(\mathbf{Q}) A_{c,v,\mathbf{k},\mathbf{Q}}^{S} & = \left( E_{c,\mathbf{k}+\mathbf{Q}}-E_{v,\mathbf{k}}\right)\delta_{(\mathbf{k},v,c),(\mathbf{k'},v',c')} A_{c,v,\mathbf{k},\mathbf{Q}}^{S} \nonumber \\ 
		& + \frac{1}{A}  \sum_{k',v',c'} W_{(\mathbf{k},v,c),(\mathbf{k'},v',c'),\mathbf{Q}} \ A_{c',v',\mathbf{k'},\mathbf{Q}}^{S}. 
	\end{align}
In the equation above, $A=A_{c} N_{k}^{2}$ is the total area of the crystal, where $A_{c}=\sqrt{3} a^{2}/2$ and $a$ is the lattice constant. Furthermore, $ W_{(\mathbf{k},v,c),(\mathbf{k'},v',c'),\mathbf{Q}}$ represents the matrix element of the many-body Coulomb potential including direct $W^d$ and exchange $W^x$ terms, and $E_{S}(\mathbf{Q})$ is the exciton energy of the $S^\mathrm{th}$ state with the center of mass momentum $\mathbf{Q}$.
	
	To simplify the calculation, we apply the Tamm-Dancoff approximation to the many-body Coulomb potential, which neglects the orbital character of the Coulomb interaction. In this case $W^{d}_{(\mathbf{k},v,c),(\mathbf{k'},v',c'),\mathbf{Q}}$ and $W^{x}_{(\mathbf{k},v,c),(\mathbf{k'},v',c'),\mathbf{Q}}$ are given by
	\begin{equation}
		W^{d}_{(\mathbf{k},v,c),(\mathbf{k'},v',c'),\mathbf{Q}}=V_{\mathbf{k}-\mathbf{k'}} \ \langle c,\mathbf{k}+\mathbf{Q}|c',\mathbf{k'}+\mathbf{Q} \rangle \ \langle v,\mathbf{k}|v',\mathbf{k'} \rangle
		\label{coulomb}
	\end{equation}
	and
	\begin{equation}
		W^{x}_{(\mathbf{k},v,c),(\mathbf{k'},v',c'),\mathbf{Q}}=V_{\mathbf{Q}} \ \langle c,\mathbf{k}+\mathbf{Q}|v,\mathbf{k} \rangle \ \langle v',\mathbf{k'}|c',\mathbf{k'}+\mathbf{Q} \rangle.
		\label{exchange}
	\end{equation}
Equation \ref{exchange} indicates that for $Q$=0, the exchange interaction is equal to zero. For a small $Q$, it is linearly proportional to $Q$. Hence for a exciton with small $Q$, we can safely neglect EXI. This allows us to calculate exciton energy spectrum and exciton wavefunctions, considering only direct Coulomb interaction term. In the absence of EXI, the ground state wave function of the exciton in valley $\tau$ is denoted by $A_{\tau,\mathbf{Q}}^{(0)}(c,v,\mathbf{k})$. 
we neglect this term.

\subsection{Electron-hole exchange interaction description}

In monolayer TMDs, the EXI strongly couples the valley  pseudospin  of  an  exciton  to  its  center-of-mass  motion. Because of the exceptionally strong Coulomb binding of excitons in monolayer TMDs, the EXI in an exciton with a finite $Q$ has to be accounted. Up to now, only direct Coulomb interaction term, described by Eq.\ref{coulomb}, is considered in exciton energy calculation. To gain a proper exciton spectrum, in the following, we incorporate the EXI into the Coulomb interactions. Taking the exchange free valley exciton doublet ($\ket{K, \mathbf{Q}}$ and $\ket{K^{\prime}, \mathbf{Q}}$) as a basis, i.e., $\ket{X} = \alpha^{K}\ket{K,Q} + \alpha^{K^\prime} \ket{K^\prime,Q}$, one can construct the matrix of exciton Hamiltonian as a $2 \times 2$ matrix:

	\begin{equation}
		\Hat{H}_{X} (\mathbf{Q}) = 
		\begin{pmatrix}
			E_{K}^{(0)}(\mathbf{Q}) + J_{KK} (\mathbf{Q}) & J_{KK^\prime}(\mathbf{Q})\\
			J^{*} _{KK^\prime} (\mathbf{Q})& E_{K^{\prime}}^{(0)}(\mathbf{Q}) + J_{KK} (\mathbf{Q}) \\
			\end{pmatrix},
			\label{hmatrix}
	\end{equation}
	with the matrix elements being given by
	\begin{equation}
		\bra{\tau, \mathbf{Q}} \Hat{H}_{X} \ket{\tau', \mathbf{Q}} = E_{\tau}^{(0)}(\mathbf{Q})\delta _{\tau, \tau'} +  J_{\tau\tau^{\prime}} (\mathbf{Q}) ,
		\label{H_matrix}
	\end{equation}
	where $E_{\tau}^{(0)}(\mathbf{Q})$ is the exciton energy in $\tau$-valley with $Q$ momentum in the absence of EXI. The wavefunctions of the exchange-free exciton states in momentum-space are described by,
	\begin{equation}
		\ket{\tau, \mathbf{Q}} = \dfrac{1}{\sqrt{\mathcal{A}}} \sum_{vc \textbf{k}} A ^{(0)}_{\tau, \mathbf{Q}} (vc \textbf{k}) \Hat{c}^{\dagger}_{c, \textbf{k} + \mathbf{Q}} \Hat{h}^{\dagger}_{v, - \textbf{k}} \ket{GS}
	\end{equation}
	with $\ket{GS}$ being the fully occupied VBs. $J_{\tau\tau^{\prime}}(\mathbf{Q})$ in Eq.\ref{H_matrix} are intra-valley ($\tau=\tau^{\prime}$) and inter-valley ($\tau \neq \tau^{\prime}$) EXIs, governed by
	\begin{align}
		J_{\tau \tau'} (\mathbf{Q})  & =   \nonumber \\
		& \dfrac{1}{\mathcal{A}} \sum _{ \substack{vc \textbf{k}\\ v'c'\textbf{k}'} }  A^{(0)}_{\tau, \mathbf{Q}} (vc \textbf{k})W^{x}_{(\mathbf{k},v,c),(\mathbf{k'},v',c'),\mathbf{Q}} A^{(0)}_{\tau', \mathbf{Q}}(v'c' \textbf{k}').
	\end{align}
For $Q\to0$, we have that $J_{K,K}(0)$=$ J_{K^\prime,K^\prime}(0)$=$J_{Q}$. Then the intra-valley and inter-valley EXI can be described by $J_{Q}\hat{\sigma}_{0}$ and $J_{Q}[\cos(2\phi)\hat{\sigma}_{x}+\sin(2\phi)\hat{\sigma}_{y}]$, respectively, where $\phi$ is the orientation  angle  of  the  center-of-mass  momentum \textbf{Q} and 
	\begin{equation}
		J_{Q}=Ry\frac{\pi}{4}\alpha^{2}|\psi(0)|^{2}\sqrt{\frac{2T_{Q}}{Ry}}   .
		\label{exchangeeq}
	\end{equation}
In Eq. \ref{exchangeeq}, $\alpha$, $T_Q$, $Ry$ and $|\psi(0)|^{2}$ are the effective fine structure, the kinetic energy of the center-of-mass motion, the Rydberg energy and the probability that an electron and a hole spatially overlap, respectively. They are defined by $\alpha=\frac{e^{2}}{\varepsilon \hbar\nu_{F}}$, $T_{Q}=\frac{\hbar^2Q^2}{2M}$, $Ry=\frac{e^{2}}{2\varepsilon a_B}$ and $|\psi(0)|^{2}\sim a_B^{-2}$\cite{dirac}, with $\varepsilon$, $a_B$, $M$ and $\nu_{F}$ being the environment-dependent dielectric constant, the Bohr radius, the  total  mass  of  the  exciton and the Fermi velocity, respectively; Recalling exciton binding energy $E_{b}=e^2/\varepsilon a_B$, the Eq. \ref{exchangeeq} can be rewritten as,
	\begin{equation}
		J_Q=\frac{\pi}{8}\frac{1}{(\hbar\nu_F)^2}\sqrt{\frac{\hbar}{2M}}\frac{E_b^3}{\sqrt{E_b}}Q   .
	\end{equation}
After calculating the elements of the $\Hat{H}_{X} (\mathbf{Q})$, we are ready to calculate the energies of an exciton and corresponding wavefunctions. Assuming that $D_{\tau,\mathbf{Q}}=E_{\tau}^{(0)}(\mathbf{Q}) + J_{\tau\tau} (\mathbf{Q})$, then we can rewrite Eq. \ref{hmatrix} in a compact form as follows,

	\begin{equation}
		\Hat{H}_{X} (\mathbf{Q}) =\Bar{E}_{+}\sigma_0+
		\begin{pmatrix}
			\Bar{E}_{-} & J_{KK^\prime}(\mathbf{Q})\\
			J^{*} _{KK^\prime} (\mathbf{Q})& -\Bar{E}_{-} \\
		\end{pmatrix},
	\end{equation}
where $\Bar{E}_{\pm}$=$(D_{K,\mathbf{Q}}\pm D_{K^{\prime},\mathbf{Q}})/2 $. Doing algebra, we finally derive the following matrix of $\Hat{H}_{X}$,
\begin{equation}
		\Hat{H}_{X} (\mathbf{Q}) =
		\begin{pmatrix}
			\Bar{E}_{+}+\Omega_{x}\cos(\theta) & \Omega_{x}\sin(\theta)e^{-i\phi}\\
			\Omega_{x}\sin(\theta)e^{i\phi}& \Bar{E}_{+}-\Omega_{x}\cos(\theta) \\
		\end{pmatrix},
		\label{comp_Ham}
	\end{equation}
where $\Omega_{x}=\sqrt{\Bar{E}_{-}^{2}+\left|J_{KK^\prime}(\mathbf{Q})\right|^{2}}$, $\theta$ and $\phi$ are the phases in Bloch sphere, which are defined by  $\theta=\cos^{-1}\left[\frac{\Bar{E}_{-}}{2\Omega_{x}}\right]$ and $\phi=-arg[J_{KK^\prime}(Q)]$, respectively. For a special case (two degenerated valley exciton states), we have $\Bar{E}_{-}$=0, which corresponds to $\theta$=$\pi/2$. Hence only intervalley EXI impacts the exciton energy. 

Diagonalizing the matrix of Hamiltoninan of $\Hat{H}_{X} (\mathbf{Q})$ in Eq. \ref{comp_Ham}, we can straightforwardly obtain the eigenvalues and eigenstates of the exciton, given by $E_{\pm}=\Bar{E}_{+} \pm \left|J_{KK^\prime}(\mathbf{Q})\right|$,
and
 	\begin{equation}
		\ket{\psi_{\pm}}=\pm e^{\pm i\phi}\sin\frac{\theta}{2}\ket{K}+\cos\frac{\theta}{2}\ket{K^\prime}   ,
	\end{equation}
where $\ket{\psi_{-}}$ ($\ket{\psi_{+}}$) represents the ground (excited) state wavefunction of the exciton. Thus the exciton dispersion splits into two well separated branches by the intervalley EXI. On the boundary of the light cone, the splitting between the two branches is estimated to be few meV. Interestingly, for this energy reference, we can rewrite the Hamiltonian in Eq. \ref{comp_Ham} as follows, $\hat{H}_{Q}^{X}=\Omega_{Q}^{X}\cdot\vec{\sigma}$, where $\vec{\sigma}=\sigma_{x}\hat{x}+\sigma_{y}\hat{y}+\sigma_{z}\hat{z}$ are the Pauli matrices representing valley pseudo-spin, and $\Omega_{Q}^{X}=\left|J_{KK^\prime}(\mathbf{Q})\right|[\sin(\theta) \cos(\phi)\hat{x}+\sin(\theta) \sin(\phi)\hat{y}+\cos(\theta) \hat{z}]$, is the effective field acting over the valley pseudospin, with $\hat{x}$, $\hat{y}$ and $\hat{z}$ being unitary vector along $x-$, $y-$ and $z-$ axis, respectively.  
In zero external magnetic field $B$, the exciton states are two fold degenerated. The presence of $B$ along $z$-direction lifts this degeneracy, giving rise to a valley Zeeman splitting as $\Delta E^{X}_{KK^\prime}=E_{K}^{X}(Q)-E_{K^\prime}^{X}(Q)$=$g_{eff}\mu_{B}B$, where $\mu_{B}$ is the Bohr magneton and effective $g$ factor $g_{eff}\simeq-2(g_{o}+2g_{s})$ with $g_{o}$ and $g_{s}$ denoting the orbital and spin $g$-factors, respectively. To visualize the orientation of valley pseudo-spin, we evaluate the expectation value of its components ${\sigma}_{x}$, ${\sigma}_{y}$ and ${\sigma}_{z}$ in the ground state. They are governed by $\langle \bar{\sigma}_{x} \rangle =-\sin\theta \cos\phi$, $\langle \bar{\sigma}_{y}\rangle =-\sin\theta \sin\phi$ and $\langle \bar{\sigma}_{z}\rangle =-\cos\theta$. Expect for a specific statement, the parameters used in our calculations are listed in Table I for the WSe$_2$ monolayer.
	\begin{table}[H]
		\centering
		\begin{tabular}{|c|c|}
			\hline
			$\hbar\nu_{F}$ (eV.\AA)  & 3.80 \\
			\hline
			$M$ ($m_0$)  & $0.46+0.42$ \\
			\hline
			$E_{b}$ (eV) & 0.27  \\
			\hline
			$E_{\tau}^{(0)}(\mathbf{Q=0})$ (eV) & 1.97  \\
			\hline
			$g_o$ & 2 \\
			\hline
			$g_s$ & 1 \\
			\hline
		\end{tabular}
		\caption{Parameters of WSe$_2$ monolayer used in our calculation. $E_{\tau}^{(0)}(\mathbf{Q=0})$ denotes the exciton energy in $\tau$ valley at $B=0$ T and $Q=0$. $m_0$ is electron mass in the vacuum.}
		\label{tab:my_label}
	\end{table}
	 
\begin{figure}[H]
\centering
\includegraphics[width=\linewidth]{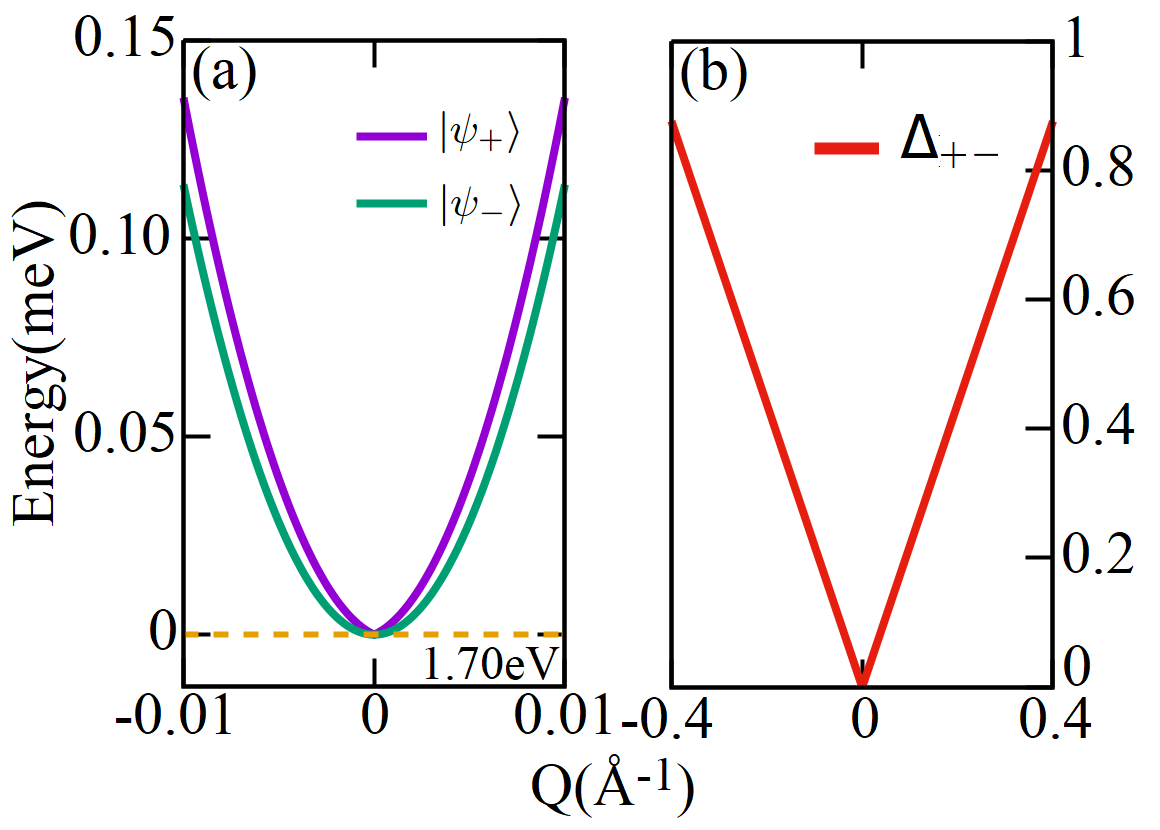}\label{energ}
\caption{(a) Exciton energies of the ground (green) and excited (violet) states in WSe$_2$ monolayer, (b) energy difference ($2J_{Q}$) between the $\ket{\psi_+}$ and $\ket{\psi_-}$ states as a function of exciton momentum $Q$, for $B=0$ T} 
\label{exchange energy}
\end{figure}

\begin{figure}[H]
\centering
\includegraphics[width=\linewidth]{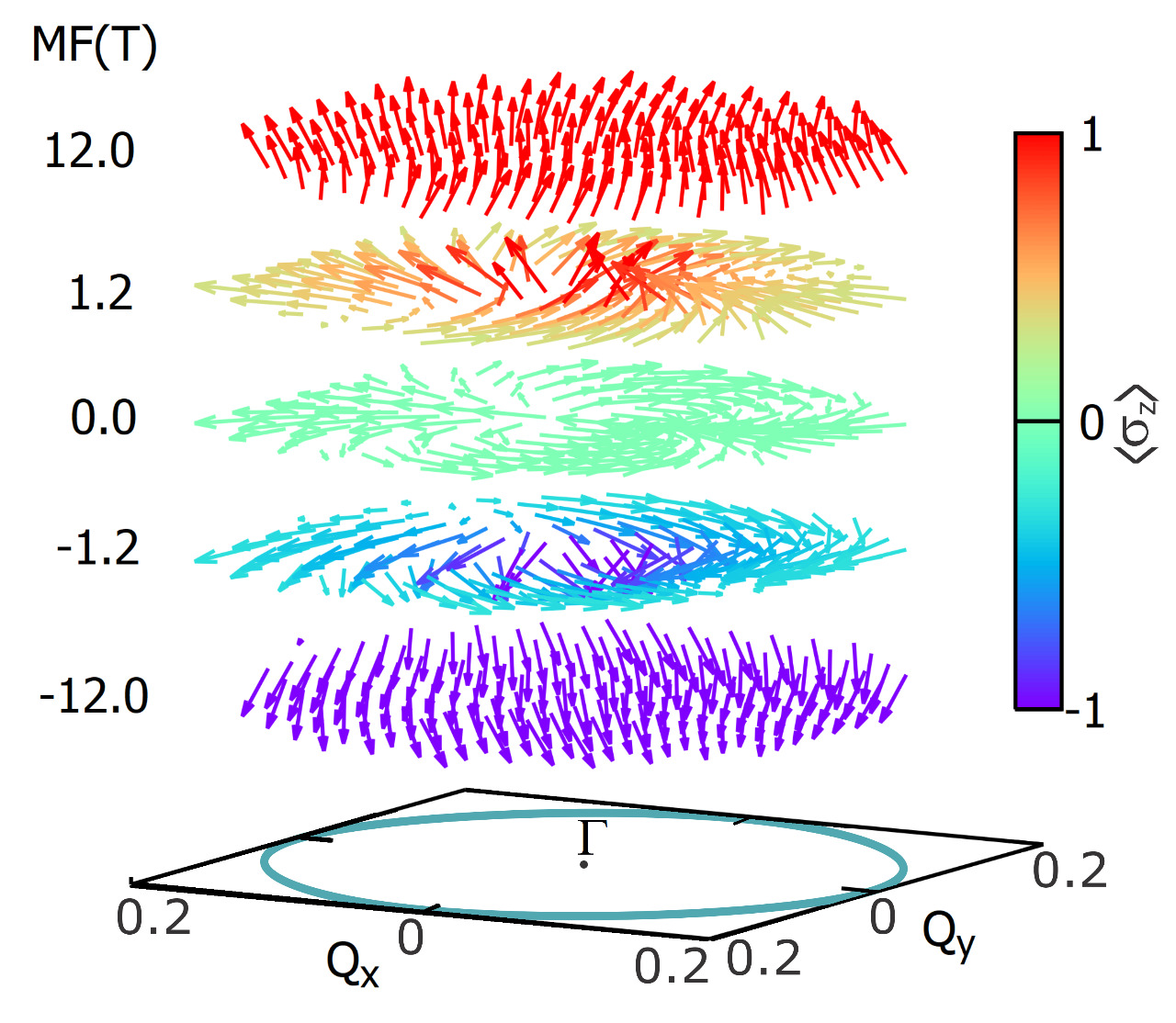}
\caption{Expectation value of valley pseudospin $\langle \bar{\sigma}_{z} \rangle$ in the ground state $\ket{\psi_-}$ around $Q=0$ for five different values of magnetic field. The color bar indicate the value of $\langle \bar{\sigma}_{z} \rangle$.} 
\label{expect}
\end{figure}

The Figure \ref{exchange energy} displays (a) exciton energies of the ground (green) and excited (violet) states and (b) energy difference ($2J_{Q}$) between the $\ket{\psi_+}$ and $\ket{\psi_-}$ states as a function of exciton momentum $Q$ in WSe$_2$ monolayer, for $B=0$ T. As expected, at $\mathbf{Q}=0$, EXI is zero. As $\mathbf{Q}$ increases from zero, it raises linearly with $Q$, obeying $\Delta_{KK'}=E_{+}-E_{-}=2J_{Q}$. Expectation value of valley pseudospin $\langle \bar{\sigma}_{z}\rangle$ in the ground state $\ket{\psi_-}$ around Q=0 for different values of magnetic field is shown in Fig. (\ref{expect}). We notice that at $B=0$, the $(x,y)$ plane pseudo-magnetic field leads to an in-plane valley pseudo-spin, whose orientation is governed by $\langle \bar{\sigma}_{z}\rangle =-\cos\theta$. With increasing external magnetic field, the valley pseudo-spin gains an out of plane components. Its orientation is determined by the summation of pseudo-magnetic field and external magnetic field. This indicates that EXI can be effectively tuned by an external magnetic field. 

\noindent

	%\begin{figure}[H]
	%	\centering
	%		\includegraphics[scale=0.07]{ef.png}
	%		\caption{(a) Effective field(colored vectors) as a function of momentum %\textbf{Q}(green dots) with color axes as the effective field in %z-direction,(b) schematic effective field lines around Q=0 with B=0 T.}
	%\label{fig3}
	%\end{figure}
	
	%\begin{figure}[H]
	%	\centering
	%		\includegraphics[scale=0.1]{texture2.png}
	%		\caption{Valley pseudospin state for the lower energy state around Q=0 %for different values of magnetic field.}
	%\label{fig2}
	%\end{figure}
	
\section{Dissipative dynamics}
The inevitable interaction of the system of interest ($S$) with the environment ($E$) gives rise to dissipative mechanisms that cause loss of coherence. Unlike a closed system, the dynamics of an open system is generally non-unitary. Within this framework, the dynamics of the reduced system $S$ is given by the time evolution of the density matrix $\rho_S$, whose equation of motion is given by the Lindblad master equation, 
\begin{equation}
\frac{\partial \rho_S}{\partial t} = -\frac{i}{\hbar}[H_S,\rho_S] + \mathcal{L}(\rho_S)
\label{Lindblad_master_equation},
\end{equation}
where the first term represents the unitary dynamics of the system driven by the hamiltonian $H_S$, the second term embraces the dissipative mechanisms written in the Born-Markov approximation. The superoperator $\mathcal{L}(\rho_S(t))$ has a specific form for each dissipative channel, the most important being those related to radiative relaxation, decoherence of quantum superpositions, and field loss mechanisms. Later, we will show the explicit forms for each of these channels.

\section{Single valley exciton qubit in dissipative TMDs}

In order to gain an insight into cavity promoted entanglement between two valley exciton qubits, let us examine  the single exciton qubit dynamics. 
The optical selection rules for interband transitions allow for valley-selective excitation at the $K$ or $K^{\prime}$ valleys using $\sigma^{+}$ or $\sigma^{-}$-polarized light, respectively. Thus, near resonance, circularly polarized excitation generates a valley exciton either in $K$- or $K^{\prime}$ valley. 
Hence in each valley, exciton can be closely described by a two-level system (TLS) where the ground state ($\ket{0}$) is related with the exciton vacuum and the excited state ($\ket{1}$) in which the exciton occupies the $1s$-like state. 
For simplicity, in the following of this section, we choose the $K$-valley exciton as representative of the system. Under the excitation of a $\sigma^{+}$ circularly right polarized light, the Hamiltonian of the system is described by two terms 
\begin{equation}
    H = H_X + H_{XL}.
    \label{single-qubit-Hamilt}
\end{equation}
The first term describes the free exciton and is given by $H_X = \frac{1}{2}\hbar\omega_{10}\sigma_z$, the second is related with the interaction of the exciton with a classical monochromatic ﬁeld whose frequency $\omega$, is close to the exciton transition frequency $\omega_{10}$. In the interaction representation and under the rotating-wave approximation (RWA), the light-matter interaction term reads,
\begin{equation}
H_{XL} = -\hbar(\Omega S_+  e^{-i\Delta t} + \Omega^\ast S_-  e^{i\Delta t}),
\label{light-matter-Hamilt}
\end{equation}
In Eq. \ref{light-matter-Hamilt}, the strength of the field-matter interaction is given by $\Omega = \frac{\mu_{10} E_0}{\hbar}$, where $\mu_{10}$ is the transition dipole moment matrix element and $E_0$ the amplitude of the laser field. Also, $\Delta = \omega - \omega_{10}$ is the detuning between the field and the exciton transition frequency. The transition operators are defined as $S_+ = \ket{1}\bra{0}$ and $S_- = \ket{0}\bra{1}$. 
%Then, in the basis of the $\ket{0}$ and $\ket{1}$, the matrix of the Hamiltonian $\Tilde{H}_{int}$ is as follows, 
%\begin{equation}
%   \Tilde{H}_{int} = -\hbar
%\begin{pmatrix}
% 0 & \Omega e^{-i\Delta t}\\
%\Omega^* e^{i\Delta t} & 0\\   
%\end{pmatrix}   .
%    \label{Matrix int  Hamiltonian}
%\end{equation}

%The dynamics of the system is given by the time evolution of the density matrix $\rho(t)$, the motion equation of $\rho$ is given by von Neumann-Lindblad master equation, 
%\begin{equation}
%\frac{\partial \rho}{\partial t} = -\frac{i}{\hbar}[H_,\rho] + \mathcal{L}(\rho)
%    \label{Lindblad_master_equation},
%\end{equation}
%where the first term on the right-hand-side yields the unitary evolution of the quantum system and the second term is the dissipative part of the evolution assuming the Born-Markov approximation. 
We assume to have two different dissipative channels, one due to the radiative decay and the other stemming from coherence relaxation. The former describing the relaxation from the excited state to the ground state is governed by the Lindblad operator $\mathcal{L}(\rho) = \frac{\Gamma}{2} (2\sigma_-{\rho}\sigma_+ -\sigma_+\sigma_-{\rho} - {\rho}\sigma_+\sigma_-)$, while the latter related to pure dephasing is dictated by $\mathcal{L}_{pd}(\rho) = \frac{\Gamma_{pd}}{2} (\sigma_z \rho \sigma_z - \rho)$. 

An analytical solution of the dissipative dynamics can be obtained by using the unitary transformation, $\Tilde{\rho}_{11} = {\rho}_{11}$, $\Tilde{\rho}_{00} = {\rho}_{00}$, $\Tilde{\rho}_{10} = {\rho}_{10}e^{-i\Delta t}$ and $\Tilde{\rho}_{01} = {\rho}_{01}e^{i\Delta t}$. Considering the initial condition $\rho_{00}(0) = 1$ and $\rho_{11}(0) = 0$, the density matrix elements are 
\begin{eqnarray*}
\rho_{11}(t) &=& \frac{R}{\Gamma + 2R}[1 - e^{-(\Gamma + 2R)t}] \\ 
\rho_{00}(t) &=& 1 - \rho_{11}(t)\\ 
\rho_{10}(t) &=& \frac{\Omega D(t)}{\Delta + i\gamma}, 
\end{eqnarray*}
where $\gamma = \frac{1}{2}\Gamma + 2\Gamma_{pd}$, $D(t) = \rho_{11}(t) - \rho_{00}(t)$, and $R = \frac{2\gamma \abs{\Omega}^2}{\Delta^2 + \gamma^2}$. For times longer than $t = (\Gamma + 2R)^{-1}$, the steady-state is reached. The stationary population of the excited state becomes
\begin{equation}
\rho_{11}^\mathrm{st} = \frac{\abs{\Omega}^2}{\frac{\Gamma}{2\gamma}(\Delta^2 + \gamma^2) + 2\abs{\Omega}^2}   .
\label{Steady}
\end{equation}
If collisions and others sources of coherence relaxation are absent, $\gamma = \frac{\Gamma}{2}$,  the population of the excited state is given by
\begin{equation}
\rho_{11}^\mathrm{st}(\Gamma_{pd}=0) = \frac{\abs{\Omega}^2}{\Delta^2 + \frac{1}{4}\Gamma^2 + 2\abs{\Omega}^2}
\label{Steady_State}   .
\end{equation}
Hence the steady-state is thus characterized by a dynamic equilibrium between stimulated transition and spontaneous decay. 

%We begin describing time evolution of exciton population after the creation of $K$-valley exciton by a circular polarized (preparation) pulse. 
\begin{figure}[H]
\centering
\includegraphics[width=\linewidth]{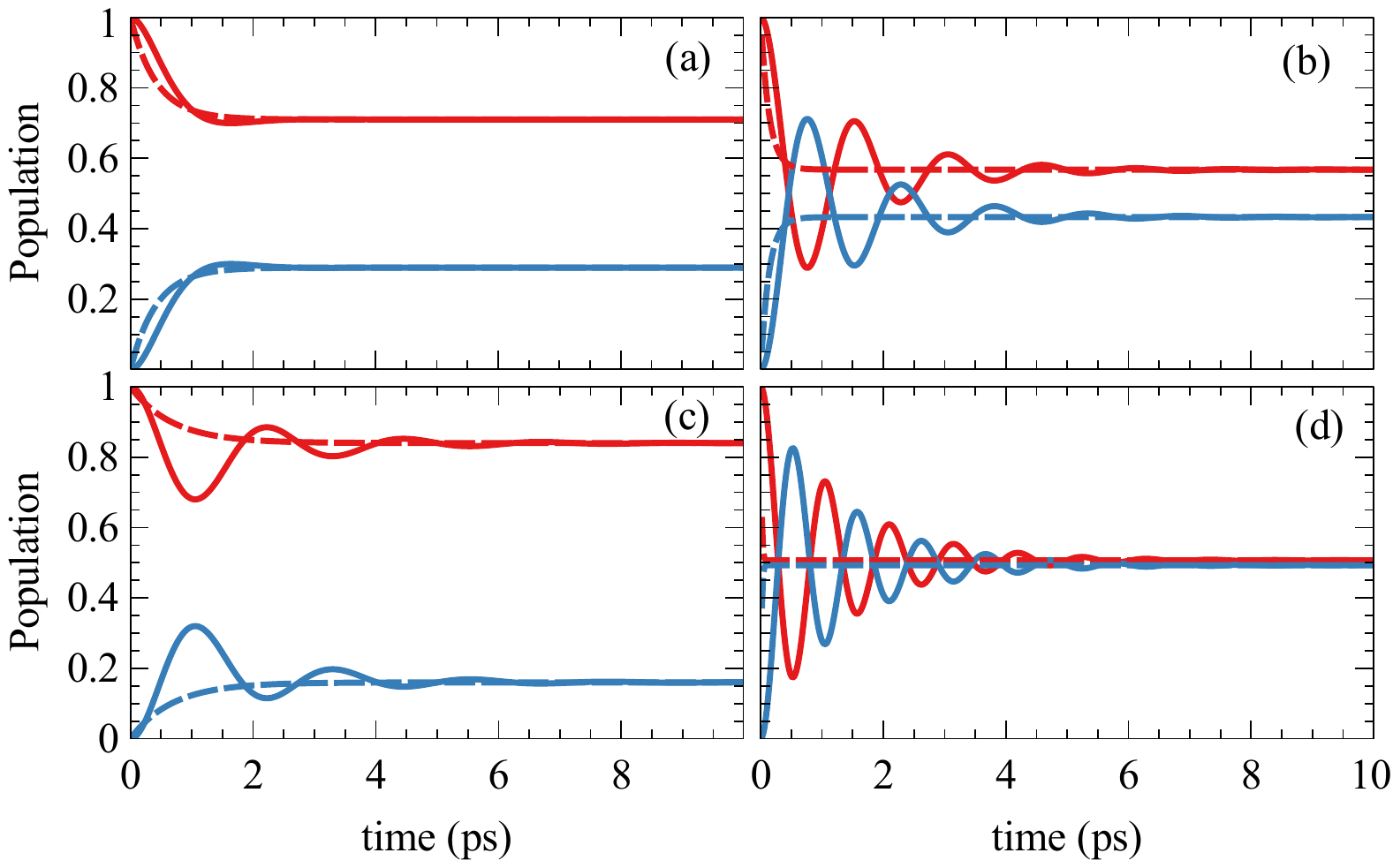}
\caption{Dynamics of single exciton population. Time evolution of the excited (blue) and ground (red) state populations of exciton for different values of the ratio $\Omega/\Delta$ and rates $\Gamma$, $\Gamma_{pd}$. (a) $\Omega/\Delta = 1$, $\Gamma = \Gamma_{pd}$ = 1 THz, (b) $\Omega/\Delta = 2$, $\Gamma$ = 1THz, $\Gamma_{pd}$ = 0 THz, (c) $\Omega/\Delta = 0.5$, $\Gamma$ = 1THz, $\Gamma_{pd}$ = 0 THz and (d) resonant case $\Delta = 0$, $\Gamma$ = 1THz, $\Gamma_{pd}$ = 0 THz. The solid and dashed lines correspond to the numerical data and analytical calculation, respectively.} 
\label{1-qubit}
\end{figure} 
\noindent
The exciton population is shown in Fig. \ref{1-qubit} for different sets of hamiltonian parameters and relaxation and dephasing rates. We notice that the numerical solution is consistent with analytical data for the large time. When the system reaches steady state, they are exact equal. It confirms the validity of our numerical method. From Eq. \ref{Steady_State}, we notice that for a large detuning, $\Delta^2 \gg  \Gamma^2, \abs{\Omega}^2$, induce a small population of the excited state $\rho_{11}\sim \Omega^2/\Delta^2$. In contrast, a near resonant strong field, $\abs{\Omega}^2 \gg \Delta^2, \Gamma^2$ produces nearly equal populations of ground and excited states $\rho_{00} \sim \rho_{11} \sim 1/2$. For the field coupling comparable to the detuning and relaxation rate $ \Omega \sim \Delta, \Gamma$, the excited state population can take appreciable values.
A simple analysis of the equations \ref{Steady} and \ref{Steady_State} allow us to show the effect on the stationary exciton population of pure dephasing processes. For near resonant condition $\Delta \sim 0$, we note that if condition $2\vert \Omega \vert ^2 + \Gamma^2/4 \gg \Gamma \Gamma_{pd}$ is satisfied, the effects of pure dephasing are negligible, that is $\rho_{11}^\mathrm{st} \sim \rho_{11}^\mathrm{st} (\Gamma_{pd}=0)$. This condition is verified in monolayer TMDs, where $\Gamma_{pd} \sim 0$ \cite{Moody2015}.

\section{Two valley exciton qubits in dissipative TMDs}
\label{section_5}
After understanding of single exciton qubit, we are ready to study the main features of two exciton dynamics.  
%and concurrence for two qubits state.
The system is compound by two excitons, located in the $K$ and $K^\prime$ valleys. The exciton in the $K$ valley ($K^\prime$) is optically driven by right (left) circularly polarized light $\sigma_+$($\sigma_-$) with frequency $\omega_R$ ($\omega_L$). EXI mediates the inter-valley coupling between the two excitons. Furthermore, an external magnetic field in the $z$ direction acts on the bright excitons inducing a Zeeman energy, $E_Z$.   
We model the system as two TLS coupled by an exchange-like interaction, for that we describe each exciton on the basis of excited and ground bare states as $\{\ket{1}_\tau, \ket{0}_\tau\}$ where $\tau = K,K^\prime$. Thus, the basis on which we represent the composite system is $\ket{1}_K\ket{1}_{K^\prime}, \ket{1}_K\ket{0}_{K^\prime}, \ket{0}_K\ket{1}_{K^\prime}, \ket{0}_K\ket{0}_{K^\prime}$, which for convenience we write in the form $\ket{11}, \ket{10}, \ket{01}, \ket{00}$. The Hamiltonian of the system can be written as 
\begin{equation}
    H = H_X +  H_{XL} + H_\mathrm{exch}, 
\label{semcavidade1}
\end{equation} 
where the first term 
\[ 
H_X = \sum_\tau \hbar \omega_e^\tau S^{\tau}_{11},  
\] 
describes the two non-interacting and unperturbed TLS, here the single exciton operators are  $S^K_{11} = \ket{1}_K\bra{1} \otimes \mathbb{1}^{K^\prime}$ and $S^{K^\prime}_{11} =  \mathbb{1}^K \otimes  \ket{1}_{K^\prime}\bra{1}$. 
Also, the energies $\hbar \omega_e^{K(K^\prime)} = E_X^{K(K^\prime)} \pm E_Z$, where $+(-)$ corresponds to $K(K^\prime)$ valley. $E_X^\tau$ are the zero-field bright $\tau$-valley exciton energy and $E_Z = \frac{\mu}{2} g_b B$ is the Zeeman energy, $\mu$ is the Bohr magneton and $g_b -4.25$ is the Lande factor for WSe$_2$ bright excitons~\cite{Foerste2020}.

The second term of (\ref{semcavidade1}) accounts for the classic exciton-laser field interaction, which in RWA is given by 
\[
H_{XL} = \frac{\Omega_+}{2} (S_-^{K}\otimes\mathbb{1}^{K^\prime}) e^{i\omega_R t} +  \frac{\Omega_-}{2} (\mathbb{1}^{K}\otimes S_-^{K^\prime})e^{i\omega_L t}+h.c,\]
where $\Omega_\pm$ is the strength of the optical coupling for $\sigma_\pm$ polarized laser assumed here as real and $h.c.$ stands for Hermitian conjugate. The third term corresponds to the contribution of the EXI 
\[
H_\mathrm{exch} = \delta (S_+^K \otimes S_-^{K^\prime} + S_-^{K} \otimes S_+^{K^\prime}), 
\]
where $\delta$ is the Coulomb exchange coupling parameter. 
An unitary rotating frame transformation is applied on the hamiltonian (\ref{semcavidade1}) designed to remove the time dependence induced by the oscillatory optical fields. The transformed hamiltonian ${H}^\prime $ is obtained from $H^\prime = UHU^\dagger + i\hbar \Dot{U}U^\dagger$, where   
\begin{eqnarray*}
U &= & e^{i(\omega_L + \frac{\omega_{RL}}{2})t} \ket{11}\bra{11} + e^{i \frac{\omega_{RL}}{2}t}\ket{10}\bra{10} + \\ 
&  &  e^{-i\frac{\omega_{RL}t}{2}} \ket{01}\bra{01} + e^{-i(\omega_L + \frac{\omega_{RL}}{2})t} \ket{00}\bra{00},
\end{eqnarray*}
and $\omega_{RL} = \omega_R - \omega_L$. The transformed Hamiltonian in the RWA approximation  is given by $H^\prime=H^\prime_X + H^\prime_{XL} + H^\prime_\mathrm{exch}$. The first term when $B=0$ is
\begin{equation}
\label{H_exc}
    H_X = \Delta_K S^K_{11} +  \Delta_{K^\prime} S^{K^\prime}_{11}, 
\end{equation}
the laser-exciton interaction is 
\begin{equation}
\label{H_laser}
     H_{XL} = \frac{\Omega_+}{2}(S^K_{-} \otimes \mathbb{1}^{K^\prime} + S^K_{+} \otimes \mathbb{1}^{K^\prime})+ \frac{\Omega_-}{2}( \mathbb{1}^{K}\otimes S^{K^\prime}_{-} + \mathbb{1}^{K}\otimes S^{K^\prime}_{+} )
\end{equation}
and EXI is 
\begin{equation}
\label{H_exch}
    H_\mathrm{exch} = \frac{\delta}{2}\left[ (S_+^K \otimes S_-^{K^\prime} )e^{-i(\Delta_K - \Delta_{K^\prime})t}  + h.c.\right].
\end{equation}
We define the energy detuning between the laser fields and excitons as $\Delta_{K} = E_{X}^K - \hbar \omega_R$, and $\Delta_{K^\prime} = E_{X}^{K^\prime} - \hbar \omega_L$, with $E_{0X}^\tau = 1.6 $eV is the zero-field exciton energy in $\tau$-valley. The experimentally accepted exchange coupling parameter for WSe$_2$ is $\delta=0.6 $meV~\cite{Molas2019,Robert2017}.

The system dynamics is contained in the density matrix which is obtained from the Lindblad master equation (\ref{Lindblad_master_equation}).
Here we consider the spontaneous decay of exciton states at the rate $\Gamma$, given by 
\begin{eqnarray}
\mathcal{L}_{K(K^\prime)}  & =\mathcal{L}_{X} + \mathcal{L}_{XX},      \label{relaxexciton}
\end{eqnarray} 
where $\mathcal{L}_{X}$ consider the recombination channels associated to one electron-hole pair in the $K$ or $K^{\prime}$ valley with a rate $\Gamma_X$ and $\mathcal{L}_{XX}$ represents the channel in which the two bright excitonic states recombine with a rate $\Gamma_{XX}$, it means that the biexciton state  $|1\rangle_K|1\rangle_{K^{\prime}}$ decays to ground state. 

Writing in the product basis, this Liouville operator represents the decay between the states $|0\rangle_K|1\rangle_{K^{\prime}}$, $|1\rangle_K|0\rangle_{K^{\prime}}$ to ground state $|0\rangle_K|0\rangle_{K^{\prime}}$ and between the states $|1\rangle_K|1\rangle_{K^{\prime}}$ to states $|0\rangle_K|1\rangle_{K^{\prime}}$ and $|1\rangle_K|0\rangle_{K^{\prime}}$. This operator is given by
\begin{eqnarray}
\label{L_X}
\mathcal{L}_{X}  & = &\frac{\Gamma_X}{2}  \left[ 2 S_-^{K(K^\prime)} \rho S_+^{K(K^\prime)} - \rho S_+^{K(K^\prime)} S_-^{K(K^\prime)} \right . \nonumber \\  
& & \left . - S_+^{K(K^\prime)} S_-^{K(K^\prime)} \rho \right]. 
\end{eqnarray} 

The description of the biexciton recombination process is provided by the following equation
\begin{eqnarray}
\label{L_XX}
\mathcal{L}_{XX} & = &\frac{\Gamma_{XX}}{2}  \left[ 2 (S_{-}^{K}\otimes  S_{-}^{K^{\prime}}) \rho (S_{+}^{K}\otimes  S_{+}^{K^{\prime}}) \right . \nonumber \\ 
& &  \left . - \rho (S_{+}^{K}\otimes  S_{+}^{K^{\prime}})(S_{-}^{K}\otimes  S_{-}^{K^{\prime}}) \right . \nonumber \\
& & \left . - (S_{+}^{K}\otimes  S_{+}^{K^{\prime}})(S_{-}^{K}\otimes  S_{-}^{K^{\prime}}) \rho \right]. 
\end{eqnarray} 

It is important to notice that the biexciton decay is significantly slower than exciton recombination, the biexciton relaxation time $\tau_{XX} = 27$ps~\cite{You2015} is large in comparison with the exciton recombination time, $\tau_X \sim 1$ ps. Thus, the biexciton relaxation rate is significantly smaller than  exciton decay rate, $\Gamma_{XX}\ll \Gamma_{X}$. Another interesting fact about dissipative dynamics in monolayer TMDs is that the mechanisms responsible for fast population relaxation do not include pure dephasing contributions \cite{Moody2015}. Thus, pure dephasing processes as elastic exciton-exciton or exciton-phonon scattering  were not considered in our calculations.

\section{Concurrence between the valley excitons}

Operationally we can define entanglement through the notion of separability: a pure bipartite state $\ket{\psi}_{AB}$ is called entangled if it cannot be separated as a direct product $\ket{\psi}_{AB}=\ket{\phi}_A \otimes \ket{\phi}_B$. In the same way, a bipartite mixed state is entangled if it cannot be represented as a mixture of factorizable the pure states, $\rho_{AB} = \sum_i p_i \ket{\alpha_i}\bra{\alpha_i} \otimes \ket{\beta_i}\bra{\beta_i}$, here $\sum_i p_i =1$ and $\ket{\alpha}$ and $\ket{\beta}$ are the pure states of subsystems $A$ and $B$, respectively.  
There are several approaches that provide a valid measure of entanglement. The choice of these measures depends on the nature of the physical system and the characteristics of the analysis to be performed.
However, the complete quantification of quantum entanglement in the general case of multiple high-dimensional q-dits is a formidable theoretical challenge that is currently under intense research.

For a simple case in which the system is composed of two exciton qubits, it is convenient to use the concurrence $\mathcal{C}(\rho)$ as a entanglement quantifier \cite{Wootters1998}. The concurrence measures the entanglement of formation of two arbitrary qubits described by the density matrix $\rho$, and is defined by 
\begin{equation}
\label{concurence}
\mathcal{C}(\rho) = \mathrm{max}(0,\sqrt{\lambda_1}-\sqrt{\lambda_2}-\sqrt{\lambda_3}-\sqrt{\lambda_4})  
\end{equation}
with $\lambda_i$ being the eigenvalues in decreasing order of
$\rho\tilde{\rho}$, where $\tilde{\rho}$ is the result of applying as the spin-flipped operation to $\rho$ (system reduced density matrix) and $\tilde{\rho}=(\sigma_{y}\otimes\sigma_{y})\rho^{*}(\sigma_{y}\otimes\sigma_{y})$. Then the Entanglement of Formation (EOF) of $\rho$ can be obtained by
\begin{equation}
\mathcal{E}(\rho) = H\left(\frac{1+\sqrt{1-C^2(\rho)}}{2}\right)  
\end{equation}
where $H(x)=-x\log_2x-(1-x)\log_2(1-x)$ with $x>0$. When $\mathcal{C} (\rho) = 0$, then a state $\rho$ is separable state. Otherwise, if $\mathcal{C}(\rho) = 1$, then it is a maximally entangled state.

\section{Results}

%\subsection{Valley exciton entanglement}

As discussed above, TMDs exhibit a rich excitonic features due to  strong Coulomb interaction. The monolayer TMDs possess a direct bandgap located at the $K$ and $K^\prime$ points in the first Brillouin zone. The time-reversal symmetry, the lack of inversion symmetry and the strong spin-orbit coupling lead to the splittings between opposite spin states and spin-valley locked band structure~\cite{Mueller2018}, which produces the chirality effect in the selection rules of the optical transitions. For example, the optical transitions can be selectively excited in the $K$ ($K^\prime$) valleys by means of right (left)-hand circularly polarized light. Optical interband transitions between two like-spin states in WSe$_2$ monolayer can generate bright valley excitons in either $K$ or ($K^\prime$) valley, see Figure \ref{Figure2}(b). Thus, optical valley selectivity together with the inter-valley coupling between two excitons in different valley induced by electron-hole exchange interaction could establish the basic architecture for applications in quantum information (QI). However, generation of entangled excitonic states which is fundamental requirement for QI applications is still a major challenge. 

 The generation of correlated photon pairs in structures formed by photonic crystal slabs and TMD monolayers has been successfully realized recently ~\cite{Wang2018,Wang2019}. Then, one may naturally speculate that an integration of monolayer TMDs in nanocavities might also yield correlated photon states ~\cite{Tokman2015}. Hence the entangled stable excitonic states might be generated by virtue of the coupling of excitons with entangled quantum optical field states of an optical cavity ~\cite{Liu2016, Li2017,Flatten2016}. \textcolor{black}{In general, the creation or preservation of excitonic entanglement depends on an interplay of characteristic of the light, detuning, inicial state and the strength of EXI. Furthermore, the loss mechanisms also play an important role.}

\subsection{Valley exciton entanglement transferred from two-mode  classical driving fields}
%\subsubsection{Valley exciton entanglement transferred from two-mode  classical driving fields}

To win an insight into generation of exciton entanglement. Let us start with our discussion about the semi-classical interaction of the composite excitonic system with circularly polarized classical field modes. 
%In this section the two-level states in each valley are coupled by continuous laser fields. 
%In the next section will be investigated the entanglement between the bright excitonic states in each valley considering a quantum field, since the system is coupled by cavity field modes. 
%The ground and excited states of the bright excitonic qubit are represented by the basis states $\{\ket{\tau, 0} ,\ket{\tau, 1} \}$, respectively, where $\tau = K, K^\prime$.

The ground and excited states of the excitonic quasiparticles in $\tau$-valley  are represented by $|i\rangle_\tau$ where $i=0,1$ and the $\tau$ is valley index referring to $K$ and $K^\prime$ valleys. Apart from the optical field, another important ingredient which can be used to tune the exciton entanglement is EXI $\delta $. It mixes the two excitonic states lying in different valleys. Namely, the excitonic states $\ket{0}_K \ket{1}_{K^\prime}$ couples with $\ket{1}_K \ket{0}_{K^\prime}$ in such a way that the non-unitary evolution can lead the system to a partially entangled state of type $\ket{\psi}^E \sim \ket{0}_K \ket{1}_{K^\prime} \pm \ket{1}_K \ket{0}_{K^\prime}$. In addition, the dynamics of the exciton states is primarily governed by the exciton-field detuning, which can be partially tuned with an external magnetic field applied along the $z$ direction. In the following, we assume that the excitonic states are driven by right (R) and left (L) hand circularlly polarized lights with frequencies $\omega_R$ and $\omega_L$. Furthermore, in order to measure the entanglement between two valley exciton qubits, we introduce the concurrence $\mathcal{C}(\rho)$ and evaluate it for two different initial states: 
a separable exciton state $\ket{\psi^X_A} = \ket{1}_K\ket{0}_{K^\prime}$ 
and a maximally entangled exciton state $\ket{\psi^X_B} = \frac{1}{\sqrt{2}} (\ket{0}_K\ket{1}_{K^\prime}-\ket{1}_K\ket{0}_{K^\prime})$. 

\begin{figure}[ht]
    \centering
    \includegraphics[scale=0.6]{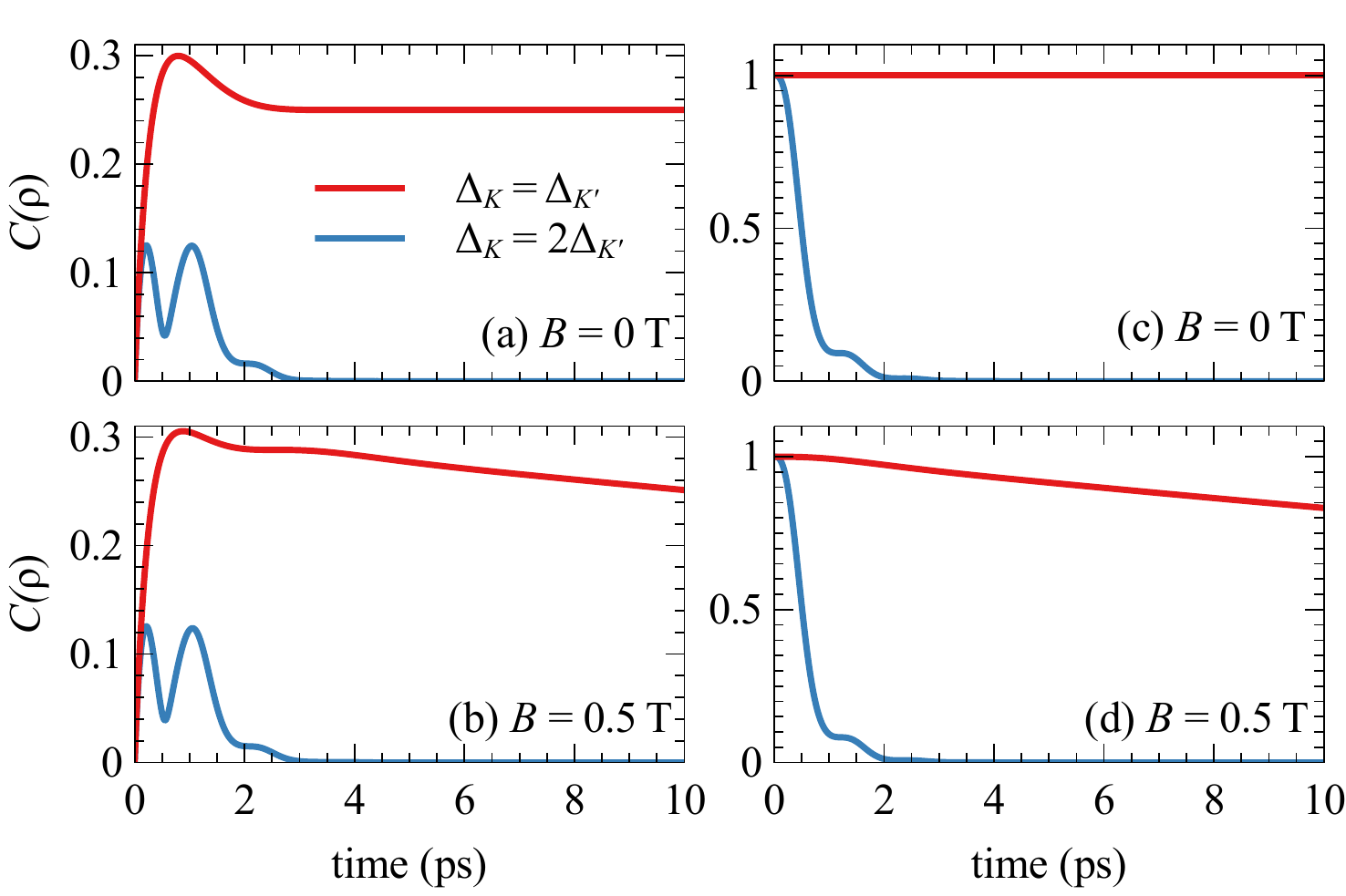}
    \caption{Dynamics of exciton concurrence under resonant detuning ($\Delta_K =\Delta_{K^\prime}= 7$meV) (red line) and  nonresonant detuning ($\Delta_K=7$meV, $\Delta_{K^\prime}=3.5$meV) (blue line). In the panels (a) and (b), the initial state is the state $\ket{\psi^X_A}$, where the exciton states are in a product state, while the panels (c) and (d) correspond to the initial state $\ket{\psi^X_B}$ in which the exciton states are written as maximally entangled state. The exciton recombination rate is $\Gamma_X = 1.0$THz and exchange coupling $\delta = \delta_0 = 0.6$meV.}
    \label{fig:Figure5}
\end{figure}

Figure~\ref{fig:Figure5} shows time-evolution of the concurrence of the exciton states driven by continuous-wave lasers. The two left panels ((a) and (b)) correspond to the concurrences of the system with exciton separable initial state ($|\Psi_A^X\rangle$), while the two right panels ((c) and (d)) are the corresponding results of the system with maximally entangled exciton initial state ($|\Psi_B^X\rangle$), in both cases $\delta_0 = 0.6$meV. 
%In the next section we are going to analyze the entanglement between the bright excitonic state when they are created by the interaction with optical modes cavity. 
%In order to compare these present results with those of the  next section, 
In each panel, there are two curves displaying the concurrence of the system with an equal exciton-field detunigs in two valleys , i.e., $\Delta_{K}=\Delta_{K^\prime}=7$meV (red curves), and with different detunings, i.e., $\Delta_{K}=2\Delta_{K^\prime}=7$meV (blue curves). Magnetic field effect is also exhibited, e.g., upper (lower) panels show the concurrence in the absence (presence) of the magnetic field ($0.5$T). It is noted that shows that in the case of $\Delta_{K}\neq\Delta_{K^\prime}$, the zero-field concurrence exhibits the same behavior with that at finite magnetic field. It drops to zero for a short time. For the system with the initial state  $|\Psi_B^X\rangle$ the concurrence falls down from one to zero rapidly. In contrast, for the system with an initial state $|\Psi_A^X\rangle$, the concurrence initially increases from zero up to $0.15$, then oscillates on the picosecond timescale and after that it goes down to zero. Although an increase of EXI (not showed here) leads to the time of the concurrence oscillation longer and the amplitude of the oscillation bigger (around  $0.8$), the entanglement still cannot achieve a stationary regime. 
On the other hand, in the case of $\Delta_{K}=\Delta_{K^\prime}$, if the initial state of the system is a maximally entangled exciton state $|\Psi_B^X\rangle$, the concurrence remains constant ($\mathcal{C}=1$). This happens because the state $|\Psi_B^X\rangle$ is an eigenstate of an exciton at $B=0$. 
Since magnetic field changes detuning via Zeeman energy, we observe that the concurrence at $B=0.5$ is no longer stationary, instead it decreases smoothly with time.  
A similar asymptotic behavior is also observed for the system with an initial state $|\Psi_A^X\rangle$. Namely, at $B=0$T, the concurrence acquires a small but stationary value $\mathcal{C}\sim 0.25$. But for $B \neq 0$, it decreases with an almost constant rate. Therefore only in the case of the detunings between the laser field energy and the transition energy of the excitons in $K$ and $K^\prime$ valleys being rigorously equal, partially entangled stationary states can be created. 
%These results also show when excitons qubits are coupled by continuous lasers field the entanglement between them, quantified by the concurrence function, is not sustained for long times if there is any detuning.   

\subsection{Generation of maximally-entangled long-lived exciton states in bimodal QED cavity}
%\subsubsection{Valley exciton entanglement in bimodal QED cavity}
We now turn our attention on to the generation of entanglement between bright excitonic states located in different valleys coupled by two modes cavity. %The sub index $\tau$ refers to $K$ and $K^\prime$ valleys.%The ground and excited states of the bright excitonic qubit are represented by the basis states $\{\ket{\tau, 0} ,\ket{\tau, 1} \}$, respectively, where $\tau = K, K^\prime$.
We consider a WSe$_2$ monolayer integrated in a bimodal optical cavity, such as a circular symmetry pillar cavity, although photonic crystal nanocavities can also be used for this purpose~\cite{JimenezOrjuela2017}. 
%An important ingredient in the problem is the exchange %interaction, which mixes the two excitonic states lying in %different valleys into a linearly polarized exciton states. %In order to control detuning between these excitonic states, %a constant magnetic field $B$ is applied along the $z$ %direction. 
The Hamiltonian of the system under the rotating wave approximation (RWA) is described by %($\hbar=1$):
\begin{equation}
    H = H_X + H_c + H_I + H_\mathrm{exch}   ,
\label{total}
\end{equation}
where the first term in Eq. \ref{total} is the exciton hamiltonian 
\begin{equation}
    H_X = \frac{\hbar}{2} \left[\omega_X S_{11}^K + \omega_X^{\prime}  S_{11}^{K^\prime}\right]   ,
\label{excitonfree}
\end{equation}
with $\hbar\omega_X = E_{0X}^K + E_Z $ and  $\hbar \omega_X^\prime = E_{0X}^{K^\prime} - E_Z$. 

The parameters associated to physical system, such as the zero-field exciton energy, are the same as those used in the previous section. 
%$\tau$ $E_{0X}^\tau = 1.6 $eV is the zero-field exciton energy in $\tau$-valley  and  $E_Z = \frac{\mu}{2}g_b B$ is the Zeeman energy, $\mu$ is the Bohr magneton and $g_b = -4.25$ is the Lande factor of the bright  excitons~\cite{Foerste2020}. In addition, $S_{11}^\tau = \ket{1_\tau} \bra{1_\tau}$ is bright exciton operator, and $\mathbb{1}^\tau$ is the identity operator in the exciton basis space. 
The bimodal cavity hamiltonian is 
\begin{equation}
H_c = \hbar\omega_a a^\dagger a + \hbar\omega_b b^\dagger b
\label{cavityfree}
\end{equation}
the cavity modes described by the bosonic operators $a$ and $b$, the mode $a$ ($b$) creates the exciton in valley $K(K^\prime)$ respectively. The each mode cavity frequencie is given by $\omega_a$ e $\omega_b$. In order to satisfy the optical selection rules, the mode $a$ ($b$) is circular right ($R$) (left ($L$))  polarized. %Their correspondent frequencies are labeled as $\omega_a = \omega_R$ and  $\omega_b= \omega_L$. 
The third term of (\ref{total}) is the exciton-cavity interaction written in dipole approximation 
\begin{equation}
    H_I = g_+ (a^\dagger S_{-}^K \otimes \mathbb{1}^{K^\prime}) +  g_- (b^\dagger \mathbb{1}^{K} \otimes S_{-}^{K^\prime}) + H.c.,
\label{interaction}
\end{equation}
where, $g_{+(-)}$ represents the coupling strength between the bright exciton in 
$K(K^{\prime})$-valleys with the mode cavity and H.c. stands for the Hermitian conjugate.
%here, $g(g^\prime)$ is the exciton-field coupling and %$S_-^\tau = \ket{0_\tau}\bra{1_\tau}$.  
%$S_{-}^{K(K^\prime)} = \ket{K,-}\bra{K, +} (\ket{K^\prime,-}\bra{K^\prime, +})$. 
Finally, the last term of (\ref{total}) refers to EXI between intervalley bright excitons  
\begin{equation}
H_\mathrm{exch} = \frac{\delta}{2} e^{-2i\theta} (S_+^{K} \otimes S_-^{K^\prime}) + \frac{\delta}{2} e^{2i\theta} (S_-^{K} \otimes S_+^{K^\prime}), 
\end{equation}
%and $S_+^\tau = \ket{1_\tau}\bra{0_\tau}$ and $\delta$ is the exchange coupling parameter which for WSe$_2$ is $\delta=0.6 $meV~\cite{Molas2019,Robert2017} and $\theta$ is the $x-y$ polar angle. 
here $\theta$ is the $x-y$ polar angle. 

%% dynamics 
The dynamics is studied by means of the density matrix operator $\rho$ of the compound system formed by the two valley bright excitons and the two-mode cavity. Its temporal evolution is given by the Lindblad master equation obtained within the Born-Markov secular approximations \ref{Lindblad_master_equation}, 
%\begin{equation}
%\frac{d\rho}{dt} = i [\rho,H] + \mathcal{L}(\rho) 
%\label{masterequation}
%\end{equation}
Here the Liouville operator $\mathcal{L} = \mathcal{L}_{K(K^\prime)} + \mathcal{L}_\mathrm{cav}$ describes the incoherent contributions to the dynamics, being $\mathcal{L}_{K(K^\prime)}$ the Lindblad operator for the exciton states (the same used to calculate the dynamics of the two exciton qubits in dissipative TMDS driven by laser fields) and $\mathcal{L}_\mathrm{cav}$ the Lindblad operator for the field cavity. 
%Here we consider the spontaneous decay of exciton %states at the rate $\Gamma$, given by 
%\begin{eqnarray}
%\mathcal{L}_{K(K^\prime)}  & = &\frac{\Gamma}{2}  \left( 2 S_-^{K(K^\prime)} \rho S_+^{K(K^\prime)} - \rho S_+^{K(K^\prime)} S_-^{K(K^\prime)} \right . \nonumber \\  
%& & \left . - S_+^{K(K^\prime)} S_-^{K(K^\prime)} \rho %\right).      
%\label{relaxexciton}
%\end{eqnarray} 
The total decay rate of the intensity of the cavity field at a rate $\kappa$ is accounted by the term  
\begin{eqnarray}
\label{relaxcavity}
\mathcal{L}_\mathrm{cav} & = & \frac{\kappa}{2}\left( 2a\rho a^\dagger - \rho a^\dagger a - a^\dagger a \rho \right) + \\ \nonumber 
& & \frac{\kappa}{2}\left( 2b\rho b^\dagger - \rho b^\dagger b - b^\dagger b \rho \right)
\end{eqnarray}

To solve the master equation of the system we construct the operators in the product state basis $\{\ket{x}_K\}\otimes\{\ket{x}_{K^{\prime}}\}\otimes \{\ket{n_a}\} \otimes \{\ket{n_b}\}$, where $\ket{x}_\tau$ are the exciton states and $\ket{n_a} (\ket{n_b})$ is the Fock basis for the right (left) polarization mode of the cavity, $\ket{x}_\tau=\ket{0}_x $ or $\ket{1}_x$ denoting vacuum and exciton state in $\tau$-valley, as described before. The  Fock state basis is numerically truncated when any relevant expected value  no longer changes when the basis size increases by one. 

%%% initial conditions  
We explore the dynamics of the dissipative system in order to analyze the dynamic behavior of quantum entanglement between excitonic qubits. For this purpose, we consider two types of initial states written as a product of exciton %$\ket{\psi_X(0)}$ 
and field 
%$\ket{\psi_f(0)}$ 
initial states, 
$\ket{\Psi(0)} = \ket{\psi_X(0)}\otimes\ket{\psi_f(0)}$: 
%(A) the field in a linearly polarized state, which corresponds to a maximally entangled Bell state~\cite{Tokman2015}, $\ket{\psi_f(0)} = \frac{1}{\sqrt{2}} \left( \ket{0}_R \ket{1}_L  + \ket{1}_R \ket{0}_L \right)$, and  the exciton state being $\ket{\psi_e(0)} = \ket{K,1} \otimes \ket{K^\prime, 0}$. (B) the field given by a Fock product state, $\ket{\psi_f(0)} = \ket{0}_R\ket{1}_L $, and  exciton states in a partial entangled state~\cite{note1} $\ket{\psi_e(0)} = C_1 \ket{K,0}\ket{K^\prime,1} + C_2 \ket{K,1}\ket{K^\prime,0}$, where $|C_1|^2 + |C_2|^2 = 1$, if $B=0$ and $E_{0X}^K=E_{0X}^{K^\prime}$, then $C_1=C_2=\frac{1}{\sqrt{2}}$.  
(i) $\ket{\Psi_A} = \ket{\psi_X}\otimes \ket{\psi_f^E}$, where the exciton subsystem is in an uncorrelated state $\ket{\psi_X} = \ket{1}_K \otimes \ket{0}_{K^\prime}$, while the field is in a linearly polarized state, which corresponds to a maximally entangled Bell state~\cite{Tokman2015}, $\ket{\psi_f^E} = \frac{1}{\sqrt{2}} \left( \ket{0}_R \ket{1}_L  + \ket{1}_R \ket{0}_L \right)$.
(ii) $\ket{\Psi_B} = \ket{\psi_X^E}\otimes \ket{\psi_f}$,   
where the excitons are in a partial entangled state~\cite{note1} $\ket{\psi_X^E} = C_1 \ket{0}_K\ket{1}_{K^\prime} + C_2 \ket{1}_K\ket{0}_{K^\prime}$, and 
the field is given by a Fock product state, $\ket{\psi_f} = \ket{0}_R\ket{1}_L $. 
Here, $|C_1|^2 + |C_2|^2 = 1$, if $B=0$ and $E_{0X}^K=E_{0X}^{K^\prime}$, then $C_1=C_2=\frac{1}{\sqrt{2}}$. 

%%% concurrence definition and some details  
%As a measure of entanglement between the two valley excitons, we adopt the Wootters concurrence, %$\mathcal{C}(\rho)$,  which is defined as~\cite{Wootters1998}
%\begin{equation}
%    \mathcal{C} = \mathrm{max}\left\{0, \sqrt{\lambda_1} - \sqrt{\lambda_2} - \sqrt{\lambda_3} - %\sqrt{\lambda_4}\right\}, 
%\label{concurrence}
%\end{equation}
%where $\lambda_i$ are the eigenvalues arranged in decreasing order of the operator $\rho(\sigma_y \otimes \sigma_y) \rho^\ast (\sigma_y \otimes \sigma_y)$, here $\rho$ is the density operator of the two exciton qubits compound system which is expressed in the two exciton basis: 
%$\ket{K,1}\ket{K^\prime,1}$, $\ket{K,1}\ket{K^\prime,0}$, $\ket{K,0}\ket{K^\prime,1}$, %$\ket{K,0}\ket{K^\prime,0}$.  
%%$\ket{1}_K\ket{1}_{K^\prime}$, $\ket{1}_K  \ket{0}_{K^\prime}$, $\ket{0}_K \ket{1}_{K^\prime}, %\ket{0}_K\ket{0}_{K^\prime}$. 
%A density operator $\rho$ of the composite system is separable if arise from product states $\rho = \rho^K\otimes \rho^{K^\prime}$, otherwise it is called an entangled state. For separable states the concurrence is $\mathcal{C} = 0$ and for maximally entangled states $\mathcal{C} = 1$.   

%\section{Results}
%% valores dos parametros espalhados no texto.

%% detuning e estado inicial 
 
As we did in the case of two exciton qubits without cavity, we define the energy detuning between the field modes and excitons, $\Delta_{K} = E_{0X}^K - \hbar \omega_a$, and $\Delta_{K^\prime} = E_{0X}^{K^\prime} - \hbar \omega_b$. If the energies of two light modes are strictly equal, then $\Delta_K = \Delta_{K^\prime}$. However, the imperfections of the cavity symmetry could modify the energy of the light modes leading to a slight difference between the detunings in valley $K$ and $K^{\prime}$, that is $\Delta_K \neq \Delta_{K^\prime}$. 
\begin{figure}[ht]
    \centering
    \includegraphics[scale=0.6]{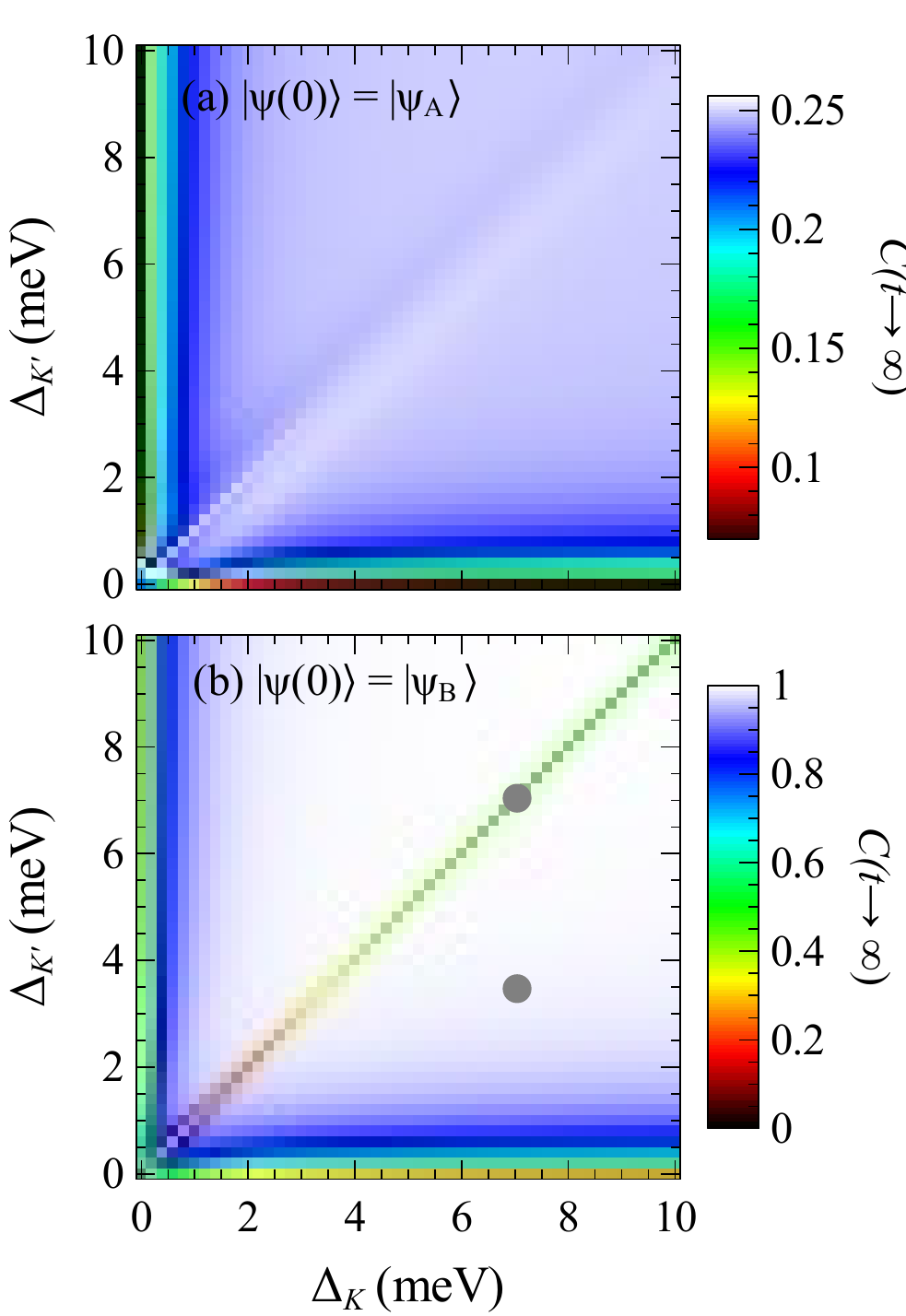}
    \caption{Asymptotic concurrence $\mathcal{C}(t \rightarrow \infty)$ as a function of exciton-field detunings when $B=0$T. Panel (a) is for the initial state $\ket{\Psi_A}$. Panel (b) is for the initial state $\ket{\Psi_B}$. (see text for a description of initial states). The following parameters are used in the calculations: EXI $\delta = \delta_0 = 0.6$meV~\cite{Molas2019,Robert2017}, exciton-cavity modes coupling $g_+ = g_- =0.1$THz, and the loss rates $\Gamma_K = \Gamma_{K^\prime} = 1$THz~\cite{De2017}, $\kappa = \kappa_0 = 0.243$THz~\cite{De2017}. The gray dots in (b) panel represent the two sets of detuning parameters $(\Delta_K, \Delta_{K^\prime})$ used in most of the calculations. }
    \label{fig:Figure1}
\end{figure}

In our system, the two exciton states are indirectly coupled via cavity field, serving as a two qubits system. The entanglement of two qubits is quantified by means of the Concurrence  ($\mathcal{C}$) which can be obtained from the exciton reduced density matrix $\rho_X$. 
This density matrix carries only the dynamics of the excitons considering the effects of all interactions contained in the total Hamiltonian. We obtain $\rho_X$ by tracing the overall system density matrix, $\rho_S$, over the cavity degrees of freedom, as~\cite{Nielsen2000} 
%The reduced density operator of the excitonic system $\rho_{X}$ is calculated by :
\begin{equation}
\rho_{X} \equiv \text{Tr}_c(\rho_S),
\label{partial_trace}
\end{equation}
where the index $c$ refers to cavity field variables. % and $\text{Tr}_c$ is the partial trace over cavity system. 
%In the excitonic basis $|0,0\rangle$, $|0,1\rangle$, $|1,0\rangle$, $|1,1\rangle$, we  derive the reduced density matrix $\rho_X$. After that, we exploit Eq. \ref{concurence} to evaluate the concurrence function ($\mathcal{C}$) .    
In order to estimate the excitonic concurrence, we must solve the Lindblad master equation, considering the full Hamiltonian (\ref{total}) and the excitonic (\ref{L_X}, \ref{L_XX}) and cavity loss mechanisms~(\ref{relaxcavity}). Once the multipartite density matrix $\rho_S(t)$ is obtained, we apply the partial trace operation (\ref{partial_trace}) to obtain the reduced density matrix of the excitonic system, $\rho_X$. After that, we evaluate concurrence $\mathcal{C}(\rho_X(t))$, which should inform us about the entanglement resulting from the multiple interactions of the excitons with the cavity quantum fields and the EXI. We also evaluate the stationary concurrence, $\mathcal{C}(t\to \infty)$, which is obtained for sufficiently long times where no temporal variation of concurrence is observed. 
 
In order to check the validity of our calculation methodology, we have performed a numerical calculation and our model prediction reproduces the result reported in Ref.~\cite{Tokman2015} (not shown here), which illustrates the dynamical emergence of maximally entangled excitonic states at well-defined times in TMD monolayers. This result is based on unitary dynamics, assuming that the initial state of the field is a Bell state ($\mathcal{C}=1$). To reproduce this result, we switched off all the scattering channels which cause the decoherence of the field and excitons. Since no decoherence mechanisms were considered, we find that the concurrence oscillates periodically and reaches its maximum value at well-defined times. This behavior indicates that, in unitary dynamics, there is an effective entanglement transfer between the quantum field and the exciton subsystem. Let us demonstrate that this transfer can be effective in non-unitary dynamics as well and leads to stationary excitonic entanglement. Unless explicitly mentioned, the following parameters are used in the calculations: EXI $\delta = \delta_0 = 0.6$meV~\cite{Molas2019,Robert2017}, exciton-cavity modes coupling $g_+ = g_- =0.1$THz, and the loss rates $\Gamma_K = \Gamma_{K^\prime} = 1$THz~\cite{De2017}, $\kappa = \kappa_0 = 0.243$THz~\cite{De2017}. 

The time-asymptotic stationary concurrence is showed in the Figure \ref{fig:Figure1}, $\mathcal{C}(t\to \infty)$ as a function of detunings $\Delta_K$ and $\Delta_{K^\prime}$ at $B=0$ for the two initial states (a) $\ket{\Psi_A}$ and (b)$\ket{\Psi_B}$. 
%The concurrence was calculated for sufficiently long times ($t>500$ps) when the time dependence on $\mathcal{C}$ induce variations less than $1\%$. 
Two interesting features are displayed. One is the impact of the initial state on the concurrence, and the other is dependence of the concurrence on detunings ($\Delta_K, \Delta_{K^\prime}$). For the system with a resonant detuning ($\Delta_K = \Delta_{K^\prime}$), if the system evolves from the initial state $\ket{\Psi_A}$, the concurrence increases initially,and  then oscillates with time, finally saturates at its maximum value of $\mathcal{C} \sim 0.3$. In stark contrast, if the system with an initial state $\ket{\Psi_B}$, the concurrence has a dip at $\Delta_K = \Delta_{K^\prime}$ and a stationary value of $\mathcal{C} \sim 0.9$ in the wide region of the figure where $\Delta_K \neq \Delta_{K^\prime}$. The underlying physics can be understood as follows. If $\Delta_K = \Delta_{K^\prime}$, then the coupling between the exciton and the cavity modes is optimal. For the system with an initial state $\ket{\Psi_A}$, the maximum entangled photon subsystem transfers part of its entanglement to the excitons, leading to an increase of exciton entanglement as well as the exciton concurrence. This behavior is more pronounced for $\Delta_K < 2 $meV. On the other side, for the system with an initially entangled exciton state $\ket{\Psi_B}$, it might transfer a fraction of its entanglement to the photon subsystem. This causes an abrupt decrease in excitonic entanglement in the region defined by $\Delta_K = \Delta_{K^\prime}$ where the exciton-field coupling is more efficient. The appreciable oscillation of the exciton concurrence on the picosecond timescale evidences the transference of entanglement between the two subsystems. In addition, the incoherent scattering channels causes an asymptotic behavior of the exciton populations, freezing the entanglement to the values of $\mathcal{C}<0.3$.  
act on the composite system
\begin{figure}[ht]
    \centering
    \includegraphics[scale=0.55]{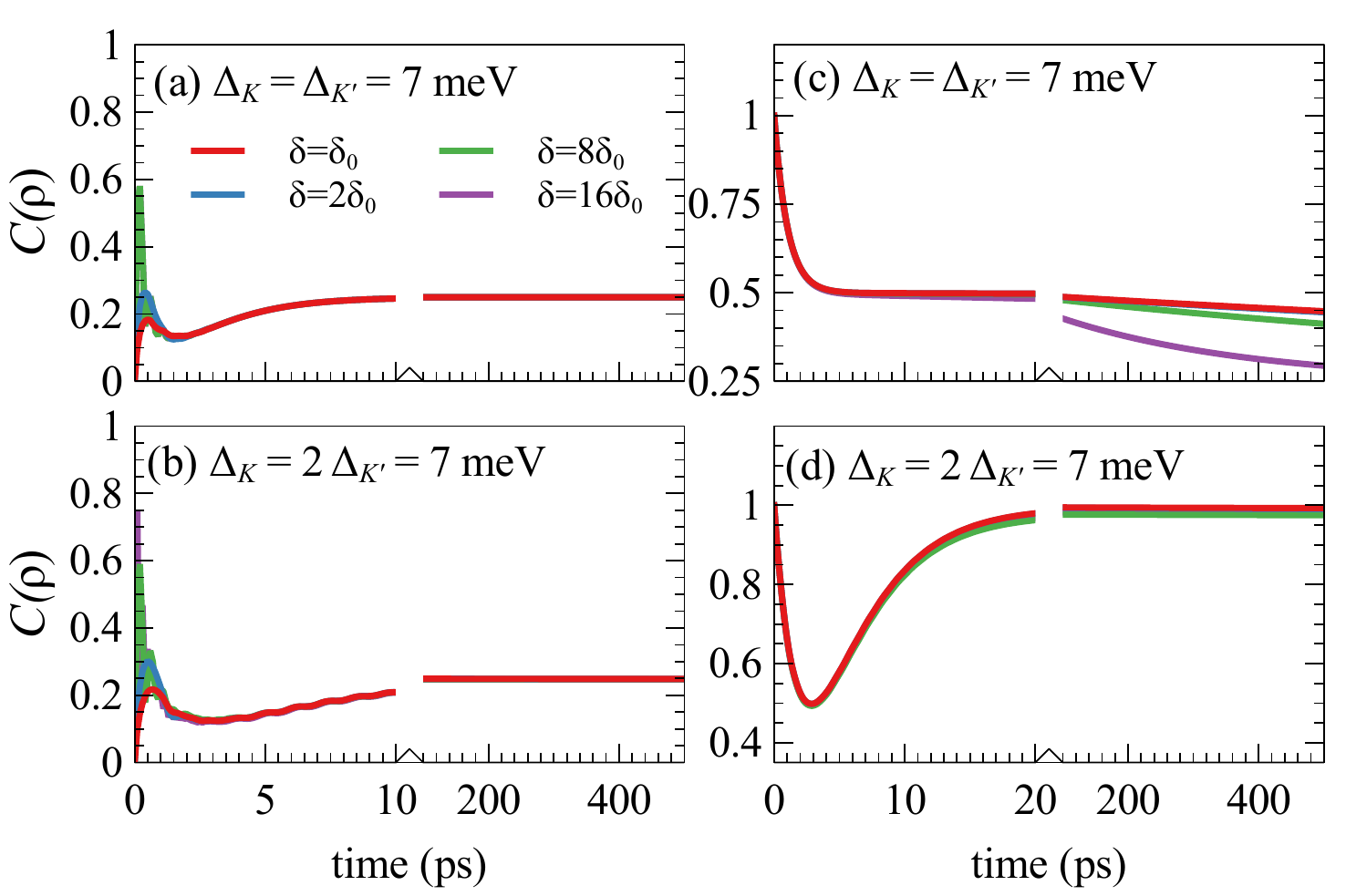}
    \caption{Time evolution of exciton concurrence at $B=0$T for different EXI strength $\delta$. Left panels (a) and (b) are for initial state $\ket{\Psi_A}$ and right panels (c) and (d) are for the system initially in $\ket{\Psi_B}$ state. Upper panels for resonant detuning ($\Delta_K = \Delta_{K^\prime}$), while the lower panels correspond to the nonresonant detuning ($\Delta_K \neq \Delta_{K^\prime}$). }
    \label{fig:Figure2}
\end{figure}

The Figure~\ref{fig:Figure2} depicts the time evolution of the concurrence of the exciton qubits for different values of the electron-hole exchange interaction strength $\delta$. The curves in the left panels ((a) and (b)) are obtained assuming that the photon entangled state $\ket{\Psi_A}$ is the initial state, while the right panels ((c) and (d)) show the concurrences with an exciton Bell state $\ket{\Psi_B}$ as an initial state. In the former (the left panels), for the systems with resonant ($\Delta_K=\Delta_{K^\prime}$) and nonresonant ($\Delta_K \neq \Delta_{K^\prime}$) detunnings, the concurrence does not exhibit a observable difference. EXI only plays a role at earlier stages of the concurrence dynamics. As time goes on, the concurrence evolves into a time independent plateau due to the decoherence. In both cases, the stationary concurrence ${C}$ is about $ 0.25$. When the initial state is switched from $\ket{\Psi_A}$ to $\ket{\Psi_B}$, however, the concurrence presents a different time-evolution and EXI dependence, see the right panels (c) and (d) of Fig.~\ref{fig:Figure2}. The entanglement of exciton qubits stemming from the initial state decays rapidly in the first few picoseconds of dynamics. This behavior depends only slightly on EXI. It is attributed to the high excitonic recombination rate. Unlike the former case shown in (a) and (b), the time evolution of the concurrence depends strongly on the detuning conditions. The behavior of the concurrence in the case shown in Fig.~\ref{fig:Figure2}(c) draws attention. Here, the initial state of the excitons corresponds to a Bell superposition of states $\vert 01 \rangle $ and $\vert 10 \rangle$. The subspace of these states behaves as a two-level system coupled by EXI. This subspace is relevant to the dynamics of the whole system since the early stages of the time evolution are governed by EXI and the initial entangled state is essentially an eigenstate of this subspace. Exciton recombination forces the concurrence to decrease rapidly to 0.5 independent of the value of the EXI coupling.
As time increases, the optical coupling comes into dynamics, the effective coupling between the excitonic system and the field now includes EXI as $\sim \sqrt{\delta^2 + g_{\pm}^2}$. Thus, by increasing the EXI strength, the effective exciton-field coupling increases and the excitonic system transfers a fraction of its entanglement to the field. %For the case of $\Delta_{K}=\Delta_{K^\prime}$ (Fig.~\ref{fig:Figure2}(c)), the maximally entangled exciton initial state quickly transfers its entanglement to the optical field. As a result, the concurrence decreases quickly. %For long times, $t>500$ ps, The concurrence decreases very slowly, for long times,  $\mathcal{C}$ reach a nearly stationary value proportional to $g/\delta$. 
For $t>5$ ps, however, the concurrence $\mathcal{C}$ decreases linearly with time with very small slope of the order of $10^{-4}$ for $\delta < 2\delta_0$. Hence one can approximately consider it as a stationary concurrence. Nevertheless, the strong EXI ($\delta > 2\delta_0$) changes this situation, suppressing the entanglement of the exciton qubits.

%For short time applications, the concurrence is practically constant at small times, $t<50$ps.
In the case of $\Delta_K \neq \Delta_{K^\prime}$, shown in panel (d), the coupling between the excitons and the cavity field is less efficient, which disfavors the entanglement transfer from the excitons to the optical field. But the interplay of this mechanism and the large excitonic recombination rate results in a pronounced decrease of exciton entanglement within the first picoseconds. Interestingly, as time goes on further, the quantum interference governed by EXI creates a superpositions of the stationary Bell-type exciton states which are eigenstates of the exciton Hamiltonian at $B=0$. Consequently, the entanglement revives and concurrence can reach a giant value close to one. It is interesting to recall that the entanglement transfer is damped by the cavity loss mechanisms.
%In the case shown in panel (d) when $\Delta_K \neq \Delta_{K^\prime}$, the exciton concurrence also reaches an asymptotic value that depends directly on  $g/\delta$, the concurrence increases when this ratio increases. 
%However, at short times the dissipative processes produce a pronounced concurrence decrease, the entanglement transferred to the field is rapidly damped due to the action of cavity losses, the exciton subsystem is driven to a steady state of almost maximum concurrence.
\begin{figure}[ht]
    \centering
    \includegraphics[scale=0.6]{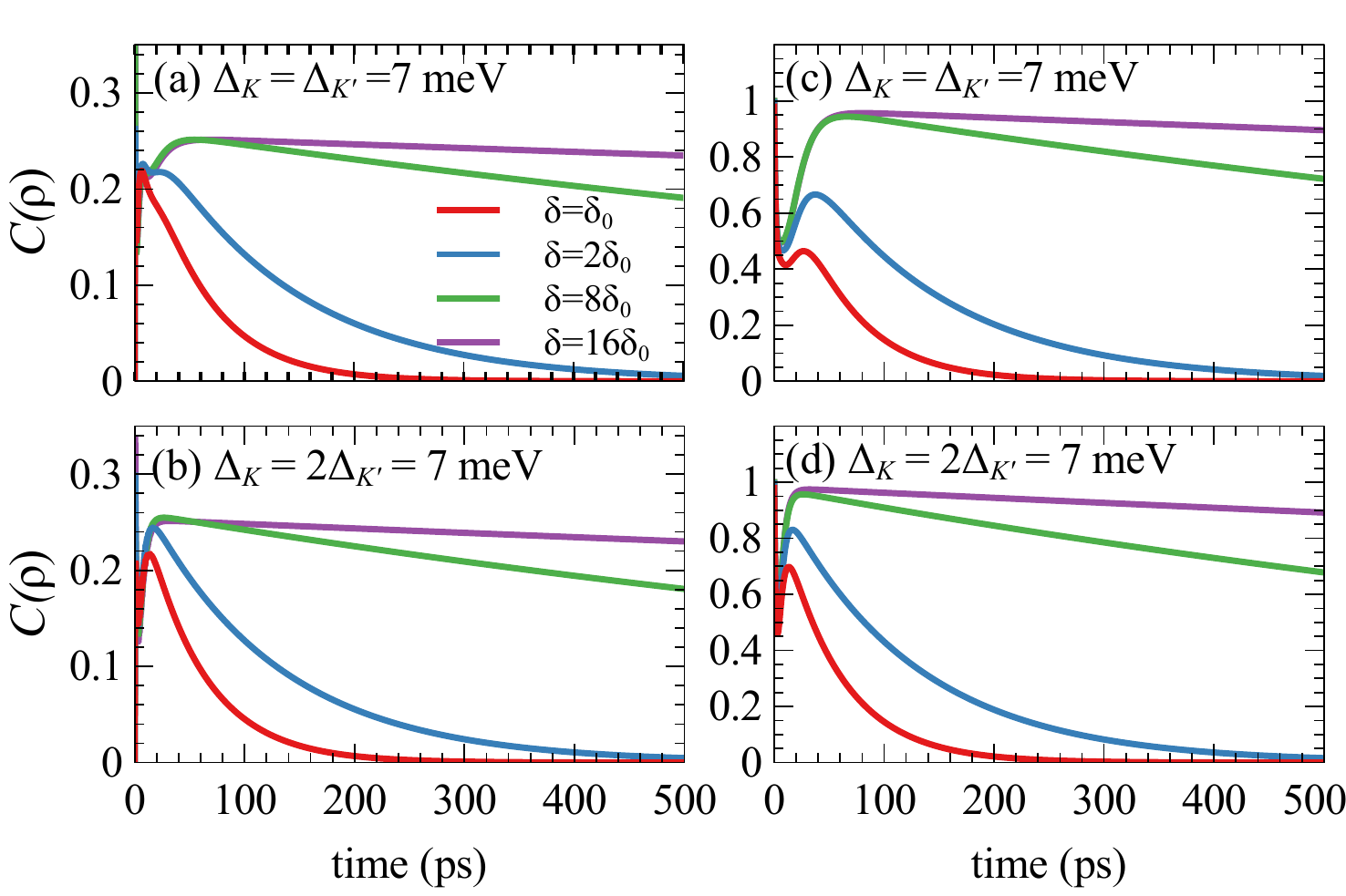}
    \caption{Same as Fig. \ref{fig:Figure2} but for $B=0.5$T. Curves shown in the left panels ((a) and (b)) are obtained with the initial state $\ket{\Psi_A}$, while the right panels ((c) and (d)) correspond to the initial state $\ket{\Psi_B}$. Upper panels for resonant detuning ($\Delta_K = \Delta_{K^\prime}$), while the lower panels correspond to the nonresonant detuning ($\Delta_K \neq \Delta_{K^\prime}$). }
    \label{fig:Figure3}
\end{figure}

Up to now, we have mainly focused our attention on the concurrence of the exciton qubits in the valley degenerate states, i.e., $E_{0X}^K = E_{0X}^{K^\prime}$, which implies that $\hbar \omega_X = \hbar \omega^\prime_X$. As known, an external magnetic field can lift the valley degeneracy of the exciton states in TMD monolayers due to Zeeman valley splitting. This allows us to study an entanglement of two non-degenerate exciton states in $K$ and $K^{\prime}$ valleys. Subjected to an external magnetic field of $B=0.5$T along the $z$-axis, it produces $0.12$ meV valley Zeeman splitting. Although this splitting is small in comparison with other energy scales of the problem, it can induce a profound impact on the concurrence behavior. Figure~\ref{fig:Figure3} shows that for a moderate value of $\delta$ ($\delta < 2\delta_0$), the lifting of exciton valley degeneracy suppresses the exciton entanglement, see the red and blue curves. In addition, this behavior occurs for an experimentally accessible EXI parameter. Because breaking exciton degeneracy makes the coherent dynamics of the exciton qubits more susceptible to incoherent effects. Then the correlation between excitons induced by a weak EXI is not strong enough to bring the system to a quasi-stationary state and maintain the entanglement. In panels (a) and (b) of Fig.~\ref{fig:Figure3}, we show time-evolution of the  exciton concurrence in the system with an initial state $\ket{\Psi_A}$. In contrast to Figs. ~\ref{fig:Figure2} (a) and (b), the concurrence is strongly dependent on exchange coupling $\delta$. 
As aforementioned, for $\delta < 2\delta_0$, the concurrence decreases exponentially. For $t>300$ps, the degree of entanglement vanishes practically. For the large value of the EXI ($\delta > 8\delta_0$), however, the concurrence can establishes its quasi-stationary value. Moreover, the behavior of the concurrence only slightly depends on the detuning $\delta_\tau$. In both resonant and non-resonant detuning cases, the asymptotic value of $\mathcal{C}\sim 0.25$ for $\delta=16\delta_0$. The right panels (c) and (d) of Fig.~\ref{fig:Figure3} show the concurrence of exciton qubits with the initial state $\ket{\Psi_B}$ for several $\delta$ values. For $\delta < 2\delta_0$, the initial excitonic entanglement is quickly dampened. The transference of entanglement to the field subsystem and the dissipative sources impede the system to reach its steady state. However, an enhancement of the excitonic correlation mediated by EXI causes a progressive recovery of the quasi-stationary behavior of the concurrence. Hence the EXI can be employed to restore exciton entanglement, independent of their detuning condition. For instance, for $\delta=16\delta_0$, the stationary value of the concurrence reaches almost as large as 1, see Fig.~\ref{fig:Figure3} (c) and (d). Since the external magnetic field lifts the degeneracy of two excitons in K and K$^{\prime}$ valleys due to the valley Zeeman splitting, the figures shown in Fig.~\ref{fig:Figure3} are the concurrence of the excitons in essentially nonresonant detuning condition. Therefore, the Fig.~\ref{fig:Figure3} (a) and (b) exhibits a similar behavior, (c) and (d) so do. In comparison of Fig.~\ref{fig:Figure2} (c) with Fig.~\ref{fig:Figure3}(c), the concurrence of two non-degenerate exciton states is quite different with that of two degenerate states.   

%However, the effect of cavity losses can be favorable to the objective of maintaining or creating excitonic entanglement. 

%% efeito do kappa 
\begin{figure}[ht]
    \centering
    \includegraphics[scale=0.6]{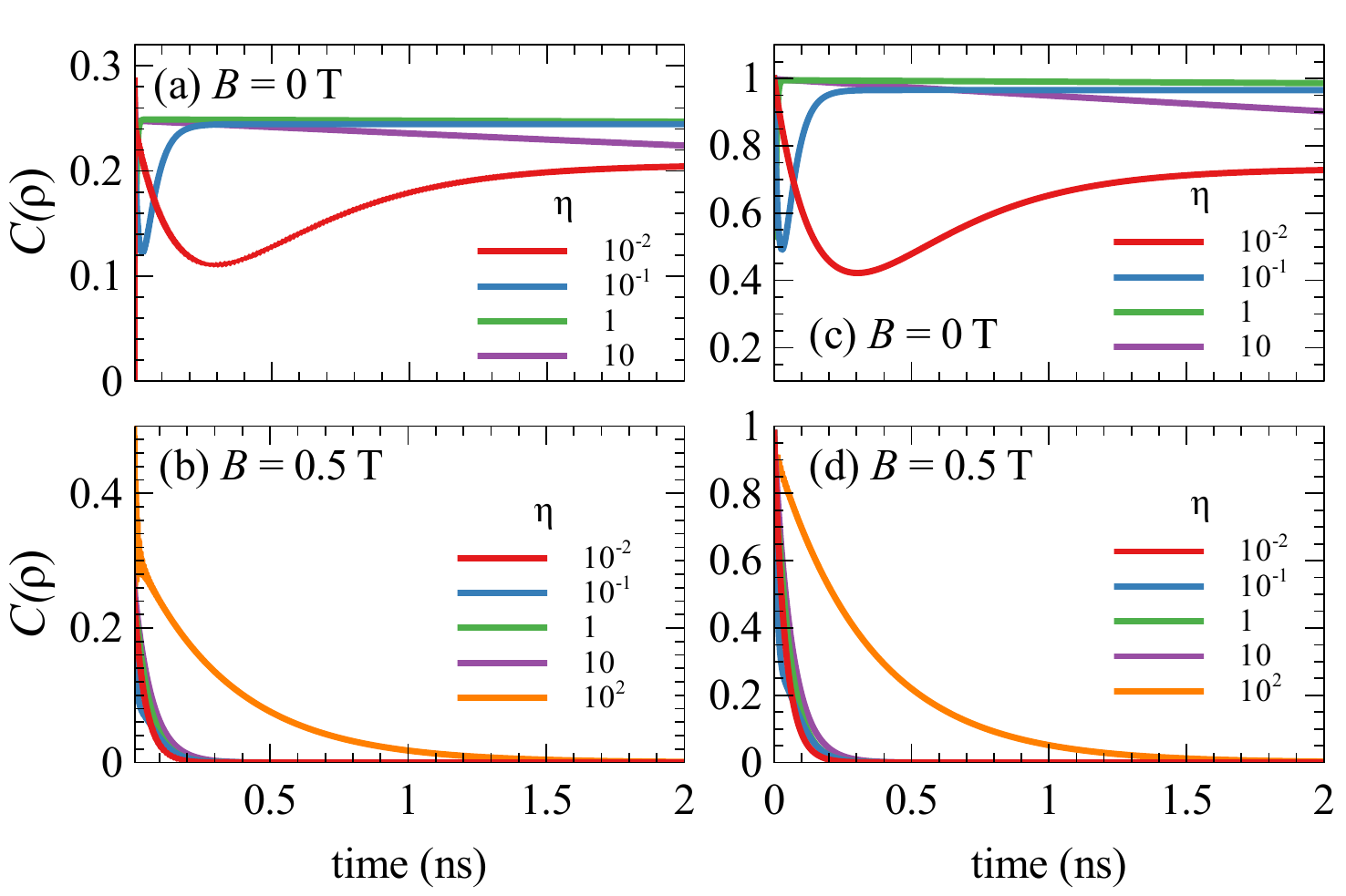}
    \caption{Temporal evolution of concurrence of two exciton qubits in WSe$_2$ monolayer integrated to the cavity in zero (upper panels) and 0.5 T (lower pannels) vertical magnetic field for five cavity loss rates, for $\Delta_K =7$meV, $\Delta_{K^\prime} = 3.5$meV, the cavity loss rate is scaled by  $\kappa = \eta \kappa_0$ with $\kappa_0 = 0.234$THz and exchange coupling $\delta = \delta_0 = 0.6$meV. Left panels (a) and (b) correspond to initial state $\ket{\Psi_A}$ and right panels (c) and (d) are obtained assuming the initial state $\ket{\Psi_B}$.}
    \label{fig:Figure4}
\end{figure}
Finally, the Figure~\ref{fig:Figure4} depicts the effects of cavity loss rate $\kappa$ on the excitonic concurrence for $\delta=\delta_0$, $\Delta_K =7$meV and $\Delta_{K^\prime} = 3.5$meV.  The results shown in the right (left) panels are obtained by assuming that the initial state is $\ket{\Psi_A}$ ($\ket{\Psi_B}$). Since the cavity field loss significantly modifies the temporal evolution of the concurrent, we use a longer timescale in this figure than that used in the previous figures. 
For convenience, $\kappa$ is scaled with an $\eta$ parameter, such that $\kappa = \eta \kappa_0$.  
In the absence of magnetic field, the two exciton states are valley degenerate states. As time goes by, the concurrence establishes its stationary value which depends only slightly on $\kappa$ in the range of approximately $(0.1 - 1) \kappa_0$. In the non-degenerate case, e.g., at $B=0.5$T, however, the concurrence inevitably diminishes to zero for sufficiently long times. In addition, the smaller the cavity loss rate, the faster the concurrence approaches to zero. 
It is interesting to recall that the rapid decay of the concurrence can be circumvented when the exchange coupling is large enough, for example, $\delta > 15 \delta_0$, see Fig.~\ref{fig:Figure3}. 
Therefore,  cavity loss plays an important role mainly in the initial stages of the concurrence dynamics, while its stationary value is only slightly altered by cavity loss.

\section{Conclusion}

We report the dissipative dynamics of two valley excitons residing in the $K$ and $K^\prime$-valleys of WSe$_2$ monolayers, coupled by EXI, under both classical and quantum optical excitations. Entanglement between valley exciton qubits has been quantitatively measured by concurrence. It depends strongly on the nature of pumping laser. Under a classical beam excitation, the concurrence is very sensitive to the status of the detuning. For the resonant detuning, stationary concurrences are obtained.  If the initial state is a maximally entangled Bell state, a storage of the concurrence is found, i.e., it remains $\sim 1$.  Otherwise, the system exhibits a small concurrence. For a non-resonant detuning, however, it is impossible to achieve a stationary entanglement. In contrast, under a quantum optical excitation, the transfer of entanglement from one subsystem (exciton/light) to the other (light/exciton) takes place. Hence a finite stationary concurrence is always found, independent of the detuning being whether resonant or non-resonant. Hence it is in strike contrast with that of classical light excitation, especially for the non-resonant detuning. In this case, while the concurrence is zero for classical beam exciton, it might be remarkably as high as 1 for quantum optical excitation. Since there is no system with a strictly resonant detuning in practice, the TMD integrated with nanocavity overcome the challenge facing by bare TMD monolayers. Finally, the lifting of the excitonic state degeneracy induced by the magnetic field suppresses stationary concurrence. Nevertheless it can be partially compensated by an increase of EXI.

\section{Acknowledgments}
This work was supported by CNPq, FAPDF, and Coordenação de Aperfeiçoamento de Pessoal de Nível Superior - Brasil (CAPES) - Finance Code 001. A. M. Alcalde and Borges H. S. acknowledge support of the Brazilian National Institute for Science and Technology of Quantum Information (INCT-IQ), grant 465469/2014-0/CNPq.

%The intrinsic, natural entanglement that the hyperfine-structure (HFS) states of the H atom store at low temperatures rapidly decreases with a growth in temperature, vanishing above a Tc threshold. An external magnetic field, however,\ref{hyperfine2021} can overcome this thermal loss of HFS entanglement. As one of the central findings of this paper, we show that an external magnetic field can induce and sustain an HFS entanglement, against all the odds of thermal effects, at temperatures well above the threshold, thus enabling magnetic-field-assisted entanglement engineering in low-temperature gases and solids.

%Reference: Natural and magnetically induced entanglement of hyperﬁne-structure states in atomic hydrogen

\bibliography{references_mod,entanglement_references}

\begin{thebibliography}{35}
\bibitem [{not()}]{note1}%
  \BibitemOpen
  \href@noop {} {}\bibinfo {note} {The coefficients are obtained from
  diagonalization of bare exciton hamiltonian $H_X=(\hbar/2) ( \omega_X^K
  \sigma_z^K \otimes \mathbb{1}^{K^\prime} + \omega_X^{K^\prime} \mathbb{1}^K
  \otimes \sigma_z^{K^\prime} )+ \delta (S^K_+ \otimes S^{K^\prime}_- + S^K_-
  \otimes S^{K^\prime}_+ ) $, the hamiltonian is block-diagonal in the basis
  space: $\{ \ket{K,0}\ket{K^ \prime,1},\ket{K,1}\ket{K^ \prime,0} \}$. We
  choose as the initial state the eigenvector $\ket{\psi_e(0)} = (C_1, C_2)^T =
  (\delta/R_-, - (\sqrt{\delta^2 + S^2}+S) /R_- )^T$ , where $S=(E_{0X}^K -
  E_{0X}^{K^\prime}) + 2E_Z$ and $R_- = \left[ 2S\sqrt{\delta^2+S^2}+
  2(\delta^2 + S^2) \right]^{1/2}$.}\BibitemShut {Stop}%
\end{thebibliography}

%%%%%%%%%% Merge with supplemental materials %%%%%%%%%%
\pagebreak
\widetext
\begin{center}
\textbf{\large Supplemental Materials: Entanglement of valley exciton qubits in transition metal dichalcogenides integrated into a nanocavity}
\end{center}
%%%%%%%%%% Merge with supplemental materials %%%%%%%%%%
%%%%%%%%%% Prefix a "S" to all equations, figures, tables and reset the counter %%%%%%%%%%
\setcounter{equation}{0}
\setcounter{figure}{0}
\setcounter{table}{0}
\setcounter{page}{1}
\setcounter{section}{0}
\makeatletter
\renewcommand{\theequation}{S\arabic{equation}}
\renewcommand{\thefigure}{S\arabic{figure}}
\renewcommand{\bibnumfmt}[1]{[S#1]}
\renewcommand{\citenumfont}[1]{S#1}
%%%%%%%%%% Prefix a "S" to all equations, figures, tables and reset the counter %%%%%%%%%%

\section{Lifetime of intralayer exciton in TMD monolayers}

The radiative lifetime featuring the stability of the exciton is an essential quantity for future application as light emitter and information storage. They are also fundamental parameters to study valley exciton dynamics.  Currently, calculations of radiative properties  typically employ simplified empirical models that can only qualitatively explain or fit the experimental data, or are carried out in the independent-particle picture, neglecting excitons altogether. While desirable, first-principles that can accurately predict exciton radiative recombination and light emission are still in their infancy. 
These approaches employ the ab initio Bethe-Salpeter equation as a starting point to compute the exciton radiative lifetimes. The radiative lifetime of excitons in monolayer TMD can be calculated using Fermi’s golden rule. We start with the usual light-matter interaction  Hamiltonian $H_{LM}$, described by,
\begin{equation}
    H_{LM} = -\frac{e}{mc}\mathbf{A}(\mathbf{r})\cdot\hat{p}   ,
\end{equation}
where $\mathbf{A}(\mathbf{r})$ is vector potential for the electromagnetic field in the Coulomb gauge, $\hat{p}$ is the momentum operator, $e$ and $m$ are electron charge and mass, respectively, and c is light speed. The vector potential can be written as $\mathbf{A}(\mathbf{r})$ = $\sqrt{\frac{2\pi\hbar{c}}{Vq}}(e^{i\mathbf{q}\cdot\mathbf{r}}\hat{\mathbf{a}}_{\lambda,\mathbf{q}}$ $\mathbf{n}_{\lambda,\mathbf{q}}+h.c.)$, where $V$ is the volume of system, $\hat{\mathbf{a}}_{\lambda\mathbf{q}}$ is photon annihilation operator, $\lambda$, $\mathbf{q}$ and $\mathbf{n}_{\lambda,\mathbf{q}}$ are the light mode, wavenumber and unit vector along light polarization direction, respectively. The lifetime ${\tau_S(\mathbf{Q})}$ for the exciton combination from an exciton state $\vert{\Psi_{S}(\mathbf{Q}), 0}\rangle$ to the final ground state $\vert{G},1_{\lambda\mathbf{q}}\rangle$ (the second term in the states gives the photon state), is given by:
\begin{equation}
    \frac{1}{\tau_S(\mathbf{Q})} = \frac{2\pi}{\hbar}\sum_{\lambda\mathbf{q}}\vert\langle{G},1_{\lambda\mathbf{q}}\vert{H_{LM}}\vert{\Psi_{S}(\mathbf{Q}), 0}\rangle\vert^2\delta(E_S(\mathbf{Q}) - E_{\lambda\mathbf{q}}),
\label{lifetime}
\end{equation}
where $E_S(\mathbf{Q})$ and $E_{\lambda\mathbf{q}}$ are the energies of the exciton and photon, respectively, $\delta$ is the delta function. The term $\langle{G},1_{\lambda\mathbf{q}}\vert{H^{LM}}\vert{\Psi_{S}(\mathbf{Q}), 0}\rangle$ can be extended as:
\begin{equation}
\begin{split}
    & \langle{G},1_{\lambda\mathbf{q}}\vert{H_{LM}}\vert{\Psi_{S}(\mathbf{Q}), 0}\rangle\\
    =& \langle{G}\vert\left[-\frac{e}{mc}\sqrt{\frac{2\pi\hbar{c}}{Vq}}e^{-i\mathbf{q}\cdot\mathbf{r}}\mathbf{n}_{\lambda,\mathbf{q}}\cdot\hat{p}\right]\sum_{c,v,\mathbf{k}} A_{c,v,\mathbf{k},\mathbf{Q}}^{S}  |c,\mathbf{k}+\mathbf{Q}\rangle \\
    &\otimes |v,\mathbf{k} \rangle \\
    \approx& -\frac{e}{m}\sqrt{\frac{2\pi\hbar}{cVq}}\mathbf{n}_{\lambda,\mathbf{q}}\cdot\sum_{c,v,\mathbf{k}}A_{c,v,\mathbf{k},\mathbf{Q}}^{S}\langle{}c,\mathbf{k}+\mathbf{Q}\vert\hat{p}\vert{}v,\mathbf{k} \rangle,
\end{split}
\end{equation}
where the long wavelength approximation, ${e}^{-i(\mathbf{k}\cdot\mathbf{r})} \approx 1$, is adopted. For a two-dimensional material assuming in the $xy$ plane, due to momentum conservation, the in-plane components of the photon wavenumber $\mathbf{q}$ = ($q_x$, $q_y$, $q_z$) and the 2D exciton momentum $\mathbf{Q}$ = ($Q_x$ , $Q_y$, 0) are equal. Since the photon momentum is small in comparison with the size of the Brillouin zone, we can safely approximate optical transition element as follows,
\begin{equation}
\langle{}c,\mathbf{k}+\mathbf{Q}\vert\hat{p}\vert{}v,\mathbf{k} \rangle \approx \langle{}c,\mathbf{k}\vert\hat{p}\vert{}v,\mathbf{k} \rangle\delta_{q_x,Q_x}\delta_{q_y,Q_y}
\end{equation}
Substituting the above matrix elements in Eq. \ref{lifetime}, simplifying the term $\sum\limits_{\mathbf{q}}$ using the two delta functions $\delta_{q_x,Q_x}$ and $\delta_{q_y,Q_y}$. The lifetime $\tau_S(\mathbf{Q})$ becomes,
\begin{equation}
\begin{split}
 \frac{1}{\tau_S(\mathbf{Q})} =& \frac{4\pi^2e^2}{m^2cV}\sum_{\lambda{q_z}}\frac{1}{q}\vert\mathbf{n}_{\lambda,\mathbf{q}}\cdot\sum_{c,v,\mathbf{k}}A_{c,v,\mathbf{k},\mathbf{Q}}^{S}\langle{}c,\mathbf{k}\vert\hat{p}\vert{}v,\mathbf{k} \rangle\vert^2\\
 &\times\delta(E_S(\mathbf{Q}) - E_{\mathbf{q}\lambda}) \\
 =& \frac{4\pi^2e^2}{m^2cV}\sum_{\lambda}\frac{L_z}{2\pi}\vert\sum_{c,v,\mathbf{k}}A_{c,v,\mathbf{k},\mathbf{Q}}^{S}\langle{}c,\mathbf{k}\vert\hat{p}_{\parallel}\vert{}v,\mathbf{k} \rangle\vert^2\\
 &\times\int_{-\infty}^{\infty}dq_z\frac{1}{q}\vert\mathbf{n}_{\mathbf{\lambda,q}}\cdot(\hat{x} + \hat{y})\vert^2\delta(E_S(\mathbf{Q}) - {{q}\hbar{c}}),
\end{split}
\label{timeint}
\end{equation}
where $\hat{x}$ and $\hat{y}$ are unit vectors along the $x$-axis and $y$-axis, respectively, $\hat{p}_{\parallel}$ is the momentum operator in an arbitrary in-plane direction, the sum on $q_z$ is transformed to an integral, employing the identity $\sum\limits_{q_z}$ = $\frac{L_z}{2\pi}\int_{-\infty}^{\infty}dq_z$ with $L_z$ being the plane-normal length ($V$ = $A_{\parallel}L_z$ is the volume, $A_{\parallel}$ being the in-plane area). With the two delta functions, we have $\mathbf{q}$ = ($Q_x$, $Q_y$, $q_z$). As $\mathbf{Q}$ = ($Q_x$, $Q_y$, 0), we get $q$ = $\sqrt{Q^2 + q_z^2}$. By introducing a unit vector $\hat{u}$ in the $x=y$ direction in $x-y$ plane, i.e., $\hat{u}$ = $(\hat{x} + \hat{y})/{\sqrt{2}}$ \cite{Grossman2015}, we can rewrite the term $\mathbf{n}_{\lambda,\mathbf{q}}\cdot(\hat{x} + \hat{y})$ = $\sqrt{2}\mathbf{n}_{\lambda},\mathbf{q}\cdot\hat{u}$. We choose two components of the polarization direction unit vector $\mathbf{n}_{\mathbf{q}}$, $\mathbf{n}_{\perp}$ perpendicular to the $\mathbf{q}-\hat{u}$ plane, and $\mathbf{n}_{\parallel}$ in the $\mathbf{q}-\hat{u}$ plane with a angle $\theta$ between $\mathbf{n}_{\parallel}$ and $\hat{u}$. Then we get $\mathbf{n}_{\parallel}\cdot\hat{u}$ = $\cos\theta$, $\mathbf{q}\cdot\hat{u}$ = $q \sin\theta$ and  $\mathbf{n}_{\perp}\cdot\hat{u}$ = 0, which means that $\mathbf{n}_{\perp}$ does not contribute to the exciton lifetime. Thus we obtain:
\begin{equation}
\begin{split}
\sum_{\lambda}\vert\mathbf{n}_{\lambda,\mathbf{q}}\cdot(\hat{x} + \hat{y})\vert^2 =& 2\vert\mathbf{n}_{\parallel}\cdot\hat{u}\vert^2\\
=& 2\cos^2\theta\\
=& 2 \left [1-(\frac{\mathbf{q}\cdot\hat{u}}{q})^2 \right ]\\
=& \frac{2}{q^2} \left[q^2-\frac{(q_x + q_y)^2}{2}\right],
\end{split}
\label{equation-sum}
\end{equation}
where we use $\mathbf{q}\cdot\hat{u}$ = $(q_x + q_y)\sqrt{2}$. Substituting Eq. \ref{equation-sum} into Eq. \ref{timeint}, we get:

\begin{equation}
\begin{split}
&\sum_{\lambda}\int_{-\infty}^{\infty}dq_z\frac{1}{q}\vert\mathbf{n}_{\lambda,\mathbf{q}}\cdot(\hat{x} + \hat{y})\vert^2\delta(E_S(\mathbf{Q}) - {{q}\hbar{c}}) \\
=& \int_{-\infty}^{\infty}dq_z\frac{2}{q^3} [q^2-\frac{(q_x + q_y)^2}{2}]\delta(E_S(\mathbf{Q}) - {{q}\hbar{c}})\\
=& \int_{-\infty}^{\infty}dq_z\frac{2q_z^2 + (Q_x - Q_y)^2}{(q_z^2 + Q^2)^{3/2}} \delta(E_S(\mathbf{Q}) - {{(q_z^2 + Q^2)^{1/2}}\hbar{c}})\\
=& \int_{-\infty}^{\infty}dq_z\frac{2q_z^2 + (Q_x - Q_y)^2}{(E_S(\mathbf{Q})\hbar{c})^{3}}\\
=& \frac{2\hbar^2c^2}{E^2_S(\mathbf{Q})}\left[2\sqrt{\frac{E^2_S(\mathbf{Q})}{\hbar^2c^2} - Q^2} +\frac{(Q_x -Q_y)^2}{\sqrt{\frac{E^2_S(\mathbf{Q})}{\hbar^2c^2} - Q^2}}\right] \equiv{} I(\mathbf{Q}) ,
\end{split}
\end{equation}
where we use $q_x = Q_x$, $q_y = Q_y$, and $q^2$ = $q_z^2 +Q^2$ in the third line; use the property of delta function in the forth line; and define this integral term as $I(\mathbf{Q})$. With Eq. \ref{wannierfunction}, we have:
\begin{equation}
\begin{split}
    & \langle{}c,\mathbf{k}\vert\hat{p}_{\parallel}\vert{}v,\mathbf{k} \rangle\\
    =& \langle{}c,\mathbf{k}\vert(\frac{\partial{H^{TB}}(\mathbf{k})}{\partial{\mathbf{k}}})\vert{}v,\mathbf{k} \rangle\\
    =& \frac{1}{\sqrt{N}}\sum_{\mathbf{R}}e^{i\mathbf{k}\cdot\mathbf{R}}\langle{W}_{c\mathbf{R}}\vert[\sum_{ij}\sum_{\mathbf{R'''}}(-i\mathbf{k}\cdot\mathbf{R}''')e^{-i\mathbf{k}\cdot\mathbf{R}'''}\\
    &t_{ij}(\mathbf{R}''')\hat{a}_{i0}^{\dagger}\hat{a}_{j\mathbf{R'''}}]
    \times\frac{1}{\sqrt{N}}\sum_{\mathbf{R}'}e^{-i\mathbf{k}\cdot\mathbf{R}'}\vert{W}_{v\mathbf{R}'}\rangle\\
    =& \frac{1}{N}\sum_{\mathbf{R}}({-i\mathbf{k}\cdot\mathbf{R}})e^{-i\mathbf{k}\cdot\mathbf{R}}t_{cv}(\mathbf{R}),
\end{split}
\end{equation}
where we use the commutation relation $\hat{p}$ = $(im_e/\hbar)$[$\mathbf{H}, \mathbf{k}$] = $\partial{\mathbf{H}}/\partial{\mathbf{k}}$. Then we get the final form for exciton lifetime:
\begin{equation}
\begin{split}
     \frac{1}{\tau_S(\mathbf{Q})} =& \frac{4\pi{}e^2}{m^2cAN^2}\vert\sum_{c,v,\mathbf{k}}A_{c,v,\mathbf{k},\mathbb{Q}}^{S}\frac{1}{N}\sum_{\mathbf{R}}({-i\mathbf{k}\cdot\mathbf{R}})e^{-i\mathbf{k}\cdot\mathbf{R}}\\
     &\times{}t_{cv}(\mathbf{R})\vert^2I(\mathbf{Q})\\
 =& \frac{4\pi{}e^2}{m^2cAN^2}{\mu_S^2(\mathbf{Q})}I(\mathbf{Q}),
\end{split}
\end{equation}
where we define the square modulus of the BSE exciton transition dipole element $\mu_S^2(\mathbf{Q})$ $\equiv$ $\vert\sum_{c,v,\mathbf{k}}A_{c,v,\mathbf{k},\mathbb{Q}}^{S}\frac{1}{N}\sum_{\mathbf{R}}({-i\mathbf{k}\cdot\mathbf{R}})e^{-i\mathbf{k}\cdot\mathbf{R}}t_{cv}(\mathbf{R})\vert^2$. In our calculation, we studied the recombination of direct excitons located within the light cone nearby the $K$ point, where $I(0)$ = $\frac{2\hbar{c}}{E_S(0)}$ and $\frac{1}{\tau_S(0)}$ = $\frac{8\pi\hbar{}e^2}{m^2E_S(0)AN^2}\mu_S^2(0)$. Considering the temperature factor, we use the relationship $\langle{\tau_{0,T}}\rangle$ = ${\tau_{0}}\cdot\frac{3M_Sc^2k_BT}{2E^2_S(0)}$ \cite{Grossman2015}, where $M_S$ is the exciton mass, $k_B$ is the Boltzmann constant and $T$ is the temperature. Then we get the results in Table \ref{table:1}.

\begin{table}[h!]
\centering
\begin{tabular}{|c |c |c |c |c|} 
 \hline
 E$_A(0)$ /eV & $M_S$ /$m_0$ & $\tau_{0,0K}$ /$ps$ & $\langle\tau_{0,4K}\rangle$ /$ps$ & $\langle\tau_{0,293K}\rangle$ /$ps$\\  
 \hline
 1.72 & 0.2 \cite{Stier2015} & 0.312 & 5.591 & 409.54 \\ 
 \hline
\end{tabular}
\caption{Radiative Lifetimes in Monolayer WSe$_2$}
\label{table:1}
\end{table}

\section{Matrix representation of Lindblad operators}
In order to simplify the notation we labeled the compound basis as: $\ket{1} = \ket{11}$, $\ket{2} = \ket{10}$, $\ket{3} = \ket{01}$, $\ket{4} = \ket{00}$. We consider the relaxation processes of the two excitonic states $K$ and $K^\prime$ to exciton vacuum state, $2 \to 4$ and $3\to 4$ at a rate $\Gamma_X$. The biexciton state relaxes to the excitonic states, $ 1 \to 2$ and $1 \to 3$, these two processes occur at the same rate $\Gamma_X$. We also consider the direct relaxation of the biexciton to vacuum state, $1 \to 4$ at a rate $\Gamma_{XX}$. 
For a general relaxation process between the states $i \to j$ , the corresponding relaxation operator is written as $L_{ij}= \sqrt{\Gamma_{ij}} \ket{j} \bra{i}$. Thus, the Liouvillian operator that describes the non-unitary dynamics induced by the relaxation mechanisms is written in Lindblad form as
\begin{equation}
    \mathcal{L}(\rho) = \sum_n \frac{1}{2}\left[ 2L_{n}\rho(t) L^{\dagger}_{n}-\rho(t) L^{\dagger}_{n}L_{n}-L^{\dagger}_{n}L_{n}\rho(t) \right ],
\end{equation}
here, the index $n$ covers all the considered $i\to j$ relaxation processes. The operator $\mathcal{L}(\rho)$ can be written in matrix form in the compound basis described above as  

\begin{equation}
\mathcal{L} = 
\begin{pmatrix}
-(\Gamma_{XX} + 2 \Gamma_X )\rho_{11} & -\frac{1}{2} \left ( \Gamma_{XX} + 3 \Gamma_X \right) \rho_{12} &  -\frac{1}{2} \left ( \Gamma_{XX} + 3 \Gamma_X \right) \rho_{13} &  -\frac{1}{2} \left ( \Gamma_{XX} + 2 \Gamma_X \right) \rho_{14}\\
 -\frac{1}{2} \left ( \Gamma_{XX} + 3 \Gamma_X \right) \rho_{21} & \Gamma_X \rho_{11}-\Gamma_X \rho_{22} & -\Gamma_{X} \rho_{23}  & -\frac{\Gamma_X}{2} \rho_{24}\\
 -\frac{1}{2} \left ( \Gamma_{XX} + 3 \Gamma_X \right) \rho_{31} & -\Gamma_X \rho_{32}  & \Gamma_{X} \rho_{11} - \Gamma_X \rho_{33} & -\frac{\Gamma_X}{2} \rho_{34}\\
-\frac{1}{2} \left ( \Gamma_{XX} + 2 \Gamma_X \right) \rho_{41} & -\frac{\Gamma_{X}}{2} \rho_{42} & -\frac{\Gamma_X}{2} \rho_{43} & \Gamma_X \rho_{33}+{\Gamma_{X}} \rho_{22}+{\Gamma_{XX}} \rho_{11}
\end{pmatrix}
\end{equation} 
Thus, the equation of motion for the density matrix, $\Dot{\rho} = -(i/\hbar)\left[H,\rho \right] + \mathcal{L}(\rho)$, can be written as a set of sixteen coupled differential equations
\begin{eqnarray*}
\Dot{\rho}_{11} &=& -\frac{i}{\hbar} \left [ \frac{g_+}{2} (\rho_{31} - \rho_{13}) + \frac{g_-}{2} (\rho_{21} - \rho_{12}) \right] - (\Gamma_{XX} + 2 \Gamma_X) \rho_{11} \\ 
\Dot{\rho}_{12} &=& -\frac{i}{\hbar} \left[ \frac{g_+}{2} (\rho_{32} - \rho_{14}) + \frac{g_-}{2}(\rho_{22} - \rho_{11}) - \delta e^{i\Delta t} \rho_{13} \right ] - \left [ \frac{1}{2}(\Gamma_{XX} +3\Gamma_X) + \frac{i}{\hbar} (\Delta_{K^\prime} - E_Z)\right] \rho_{12} \\
\Dot{\rho}_{13} &=& -\frac{i}{\hbar} \left [ \frac{g_+}{2} (\rho_{33} - \rho_{11}) + \frac{g_-}{2}(\rho_{23} - \rho_{14}) - \delta e^{-i\Delta t} \rho_{12} \right ]
- \left [ \frac{1}{2}(\Gamma_{XX} +3\Gamma_X) + \frac{i}{\hbar} (\Delta_K + E_Z)\right] \rho_{13} \\
\Dot{\rho}_{14} &=& -\frac{i}{\hbar} \left[ \frac{g_+}{2} (\rho_{34} - \rho_{12}) + \frac{g_-}{2}(\rho_{24} - \rho_{13})\right] - \left[ \frac{1}{2}(\Gamma_{XX} + 2\Gamma_X) + \frac{i}{\hbar} (\Delta_K + \Delta_{K^\prime}) \right]\rho_{14} \\
\Dot{\rho}_{22} &=& -\frac{i}{\hbar} \left [ \frac{g_+}{2} (\rho_{42} - \rho_{24}) + \frac{g_-}{2}(\rho_{12} - \rho_{21})  + \delta (e^{-i\Delta t} \rho_{32} - e^{i\Delta t} \rho_{23} )\right]  + \Gamma_X \rho_{11} - \Gamma_X \rho_{22} \\
\Dot{\rho}_{23} &=& -\frac{i}{\hbar} \left [ \frac{g_+}{2} (\rho_{43} - \rho_{21}) + \frac{g_-}{2}(\rho_{13} - \rho_{24}) + \delta e^{-i\Delta t} (\rho_{33} - \rho_{22})\right] - \left[ \Gamma_X + \frac{i}{\hbar} (\Delta_K - \Delta_{K^\prime} + 2E_Z)\right ]\rho_{23} \\
\Dot{\rho}_{24} &=& -\frac{i}{\hbar} \left [ \frac{g_+}{2} (\rho_{44} - \rho_{22}) + \frac{g_-}{2}(\rho_{14} - \rho_{23}) + \delta e^{-i\Delta t} \rho_{34} \right] - \left[ \frac{\Gamma_X}{2} + \frac{i}{\hbar} (\Delta_K + E_Z) \right] \rho_{24} \\
\Dot{\rho}_{33} &=& -\frac{i}{\hbar} \left [ \frac{g_+}{2} (\rho_{13} - \rho_{31}) + \frac{g_-}{2}(\rho_{43} - \rho_{34})  + \delta (e^{i\Delta t} \rho_{23} - e^{-i\Delta t} \rho_{32} )  \right] - \Gamma_X \rho_{33} + \Gamma_X \rho_{11} \\
\Dot{\rho}_{34} &=& -\frac{i}{\hbar} \left [ \frac{g_+}{2} (\rho_{14} - \rho_{32}) + \frac{g_-}{2}(\rho_{44} - \rho_{33}) + \delta e^{i\Delta t} \rho_{24} \right] - \left[ \frac{\Gamma_X}{2} + \frac{i}{\hbar} (\Delta_{K^\prime} - E_Z) \right] \rho_{34} \\
\Dot{\rho}_{44} &=& -\frac{i}{\hbar} \left [ \frac{g_+}{2} (\rho_{24} - \rho_{42}) + \frac{g_-}{2}(\rho_{34} - \rho_{43})  \right] + \Gamma_X \rho_{33} + \Gamma_X \rho_{22} + \Gamma_{XX} \rho_{11},  
\end{eqnarray*}
where $\Delta = \Delta_K - \Delta_{K^\prime}$  and  $\rho_{ji} = \rho_{ij}^\ast$. The solutions of the system of equations must satisfy the constraint $\sum_{ii}\rho_{ii}=1$.

\end{document}